\documentclass[fleqn,10pt]{wlscirep}
\usepackage[utf8]{inputenc}
\usepackage[T1]{fontenc}
\usepackage{soul}
\usepackage{ulem}
 \usepackage{graphicx}
\usepackage{subfigure}
\usepackage{wrapfig}

\title{
 Transport of pseudothermal photons  through an anharmonic   cavity
}

\author[1,2,3,4,*]{Dmitriy S. Shapiro
} 
	\affil[1]{Dukhov Research Institute of Automatics (VNIIA),   127055 Moscow, Russia}
	\affil[2]{	Department of Physics, National Research University Higher School of Economics, 101000 Moscow, Russia}
 \affil[3]{	Laboratory of Superconducting Metamaterials, National University of Science and Technology MISiS,   119049 Moscow, Russia}  
\affil[4]{	V. A. Kotel'nikov Institute of Radio Engineering and Electronics, Russian Academy of Sciences, Moscow 125009, Russia}
\affil[*]{shapiro.dima@gmail.com}

\begin{abstract}

Under nonequilibrium conditions, quantum optical systems reveal unusual properties   that might be distinct from  
  those  in condensed matter.
 The fundamental reason is that  photonic eigenstates   
can have 
arbitrary occupation  numbers, whereas in electronic systems these are limited by the Pauli principle.
Here, we address   the steady-state transport of pseudothermal photons between two   waveguides connected through  
a cavity with  
Bose-Hubbard  
interaction  between photons. One of the waveguides is  subjected to a broadband incoherent  pumping. We predict a continuous transition between the regimes of  Lorentzian  and Gaussian chaotic light emitted by the cavity. 
 The rich variety of nonequilibrium transport regimes is revealed by the zero-frequency noise. There are three limiting cases, in which the noise-current relation is characterized by a power-law, $S\propto J^\gamma$. 
 The Lorentzian   light         corresponds  to  
Breit-Wigner-like transmission and   
$\gamma=2$. The Gaussian     regime   corresponds to  many-body transport with  the   shot noise ($\gamma=1$) at large currents; at low currents, however,  we find  an unconventional    exponent $\gamma=3/2$ indicating  a nontrivial  interplay between multi-photon transitions and incoherent pumping.
 The nonperturbative solution for photon  
 dephasing is obtained in the framework of the Keldysh field theory and Caldeira-Leggett effective action. 
  These findings might be relevant   for  experiments on photon blockade in superconducting qubits,  
    thermal states transfer, and        photon statistics probing.
    
 \end{abstract}

\begin{document}
\flushbottom
\maketitle

\section*{Introduction}
Experimental and heoretical     studies of  
  out of equilibrium   cavity and circuit  QED
 have shown remarkable progress during the last   decade  \cite{Haroche:2020aa, Blais:2020aa, Frisk-Kockum:2019aa,kirton2018introduction,Miller_2005}. 
 This research area  covers a diverse class of driven-dissipative phenomena and  quantum phase transitions~\cite{Baumann, PhysRevLett.121.040503,Magazzu:2018aa,Buchhold:2013aa}.
  There,   observation of  photon-photon correlations and quantum state transfer  
 becomes possible in hybrid systems~\cite{Clerk:2020aa,PhysRevLett.107.220501, Srinivasan2011,macha2014implementation,braumuller2017analog} where transmission lines are coupled to  nonlinear quantum oscillators, such as superconducting transmon qubits or anharmonic cavities.
 In particular, the photon-photon interaction can be probed in  the  photon blockade effect as suggested in Ref.\cite{PhysRevLett.79.1467}, an optical counterpart of the Coulomb blockade in electronic devices. This is an intriguing phenomenon where a driven quantum anharmonic  oscillator    emits     anticorrelated   photon 'trains'    indicating   their sub-Poissonian statistics~\cite{birnbaum2005photon,PhysRevLett.107.053602,Photon_Blockade_corr}.  
  
 In this work, we explore the transport of incoherent photons  
 through the cavity with   anharmonicity (Kerr energy)    smaller than the excitation frequency and bandwidth of the input signal.     This   is a bosonic counterpart of thermal or   voltage biased electronic  level with Coulomb interaction~\cite{PhysRevB.50.5528,Giazotto:2006aa}.
  Our findings are motivated mostly by experiments    on  photons statistics~\cite{Dmitriev:2019aa,Honigl-Decrinis:2020aa,Zhou:2020aa} and      thermal   states propagation\cite{Goetz:2017aa} that are 
  relevant for various applications such as   a microwave    non-classical light emission~\cite{PhysRevLett.119.137001}, thermometry~\cite{PhysRevLett.119.090603}, and quantum states transfer~\cite{PhysRevA.87.022306}.
  
  The moderate anharmonicity does not lead to a well-developed photon blockade, {\it i.e.}, photon number in the cavity is proportional to the input drive and  there is no saturation of the photon current. Nevertheless, it is known that even weak anharmonicity results in the antibunching of photons and their negative correlations\cite{Liew:2010aa,Bamba:2011aa}. In our studies, we find that wideband incoherent pumping induces bunching of photons. We predict a transition from    Lorentzian  to Gaussian pseudothermal light, as follows from second-order  intensity correlator $g^{(2)}$.  
 However, the emitted light   possesses a partial coherence. This results in intriguing behavior in the  integral characteristics of the emitted (transmitted) photons such as their zero-frequency noise, written $S$.  Contrary to $g^{(2)}$ that  resolves short timescales, $S$  is provided by photon counting during  long time intervals.   
  The unusual noise-current relations derived  here represent an interplay between photon-photon interaction and strong incoherent    drive which brings the system far from equilibrium.

  The aim of this work    is twofold. First, we analyze    nonequilibrium noise, Fano factor, decoherence, and transmission spectrum of the cavity photons.   We consider   a wide range of  behaviors ranging from      the single-particle transfer  with Breit-Wigner transmission    to a many-body regime with unconventional shot noise.  
   The second purpose is methodological. We  apply the Keldysh path integral technique \cite{kamenev2011field,Altland:2010aa}, and adopt the concept of dissipative Caldeira-Leggett action\cite{Caldeira:1981aa,Shnirman:2016aa} to a   driven oscillator with Bose-Hubbard interaction.
   Although these methods found their  application  in   condensed matter theory a long time ago,  in the cavity and circuit  QED,  they have started to gain attention much later. Nowadays, applications of Keldysh formalism  to quantum optics  is an    active area of research  ~\cite{Maghrebi:2016aa,PhysRevResearch.2.033196,dalla2013keldysh,kirton2018introduction,PhysRevLett.110.195301}.  In our methodology, a nonperturbative solution for quasi-classical fluctuations, which takes into account many-body effects and decoherence
    of     transmitted photons is proposed.

\begin{wrapfigure}{r}{0.5\textwidth}
   \centering	\subfigure[Setup\label{setup}]{\hspace{-4pt}	\includegraphics[width=0.49\textwidth]{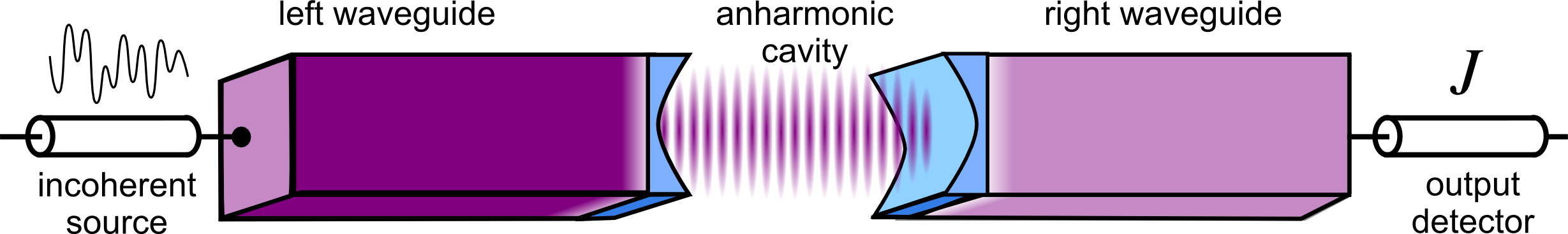}} \\ 
   \subfigure[Spectrum. Incident and outgoing distribution functions \label{spectrum}]{\includegraphics[width=0.37\textwidth]{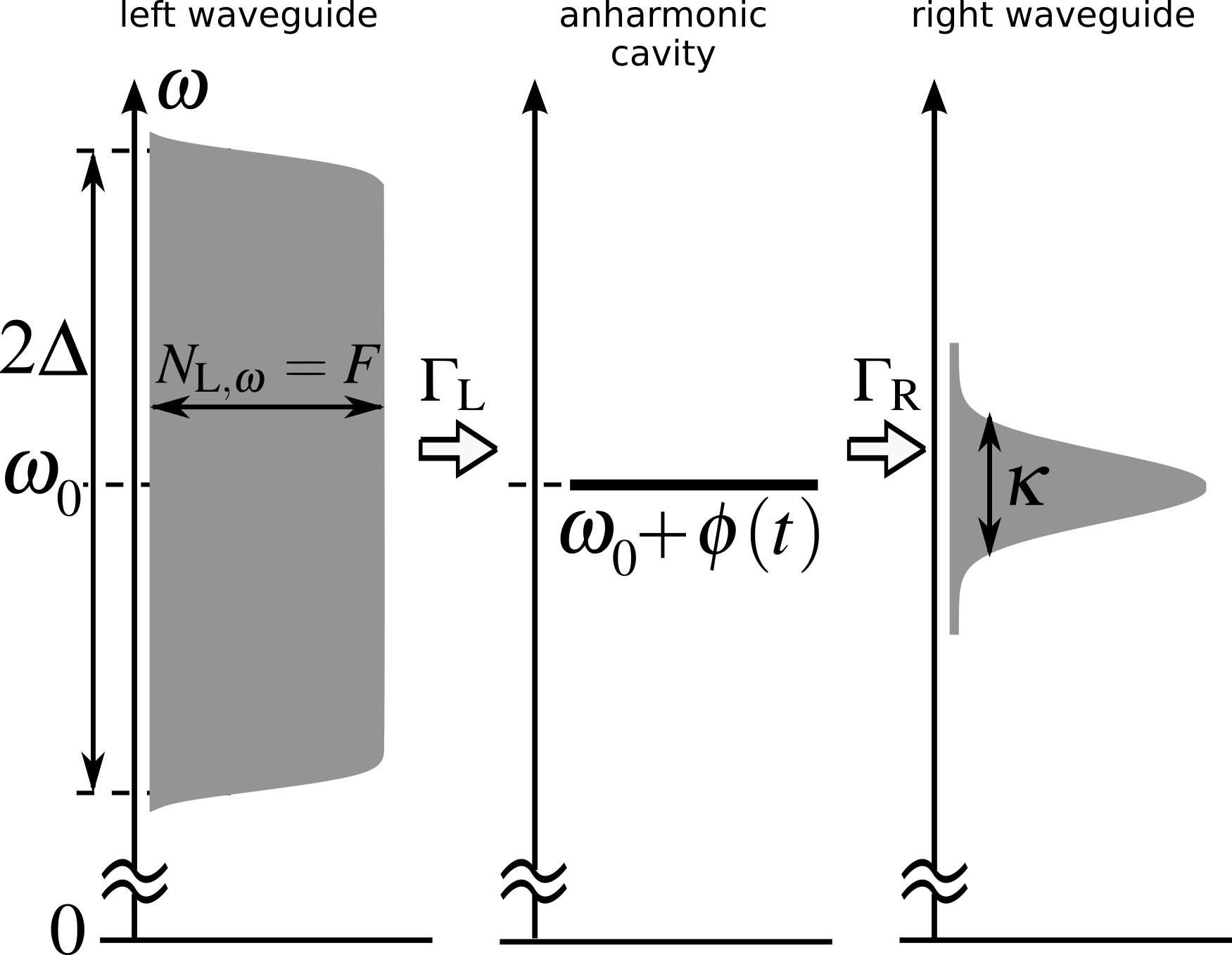}}
   	\caption{
   		(a) Sketch of the setup. The incoherent drive source emits   photons into the left waveguide. The difference in  filling color transparencies stands for a difference in    light intensities.   The standing wave between      blue mirrors is  the cavity  mode.
   		The current $J$ is measured in the output detector.    		(b)~The spectrum of the incident and outgoing photon states. Incident photons have flat spectrum $N_{\rm L,\omega}=F$ of the bandwidth $\Delta$ (gray color),   nonequilibrium outgoing   distribution function have Gaussian shape of width $\kappa$.  Fluctuating   potential $\phi(t)$ is induced by the Bose-Hubbard interaction.
   } \vspace{10pt}  \end{wrapfigure}

 The sketch of the hybrid system under consideration is shown in Fig.~\ref{setup}; it is implemented as two open waveguides  
   coupled through an  anharmonic  cavity\cite{PhysRevLett.79.1467}. 
    Alternatively, this can be a circuit QED setup, similar to that reported in Ref.~\cite{Gambetta:2006aa}, where a multilevel  transmon  qubit is coupled to superconducting transmission lines.

    In Fig.~\ref{setup}, the left waveguide is connected to a source of thermal     photons, while outgoing light is measured by a photon detector in the right waveguide.  The distribution function of incident states  in the left waveguide  
    is assumed flat $N_{{\rm L},\omega}= F  $ in the frequency range $\omega\in [\omega_0-\Delta; \ \omega_0+\Delta]$ and zero otherwise, as shown   in  Fig.~\ref{spectrum}; $\omega_0$ is the cavity mode frequency and $\Delta$ is the source linewidth. The outgoing states in the right waveguide can have Lorentzian or  Gaussian nonequilibrium distribution functions. The Gaussian distribution is determined by a width $\kappa$  depending nontrivially on $F$.
    
    The  number of photons  $F$   in a particular incident mode  can be less or greater than unity.   We assume the following ordering of the relevant frequencies
   \begin{equation} 
  	\omega_0\gg\Delta\gg  
	\epsilon,  \Gamma_{\rm L,R}  , \kappa
	\ , \label{cond}
  \end{equation}
     where  
     $\epsilon$ is the anharmonicity,  $\Gamma_{\rm L}$ and $\Gamma_{\rm R}$ are  relaxation rates due to the  coupling with the  left and the  right waveguides, and their sum    $\Gamma=\Gamma_{\rm L}+\Gamma_{\rm R}$ determines the  bare relaxation rate. (We set Planck constant $\hbar=1$ hereafter.)  
      The condition (\ref{cond}) is motivated by typical parameters of   nonlinear optical cavities~\cite{birnbaum2005photon} where the anharmonicity is induced by a coupling of the cavity mode with a trapped atom. Also, this condition is accessible in superconducting systems based on transmon qubits coupled to transmission lines~\cite{braumuller2017analog,PhysRevLett.107.053602,Photon_Blockade_corr}. 
     According to (\ref{cond}), antiresonant processes  (counter-rotating wave)  are neglected and the quantum    dynamics is determined by the following  $U(1)$   Hamiltonian with continuous symmetry,   \begin{equation}
   	\hat H= \hat H_{\rm ac}+\hat H_{\rm L}+\hat H_{\rm R}+\hat H_{\rm tL}+\hat H_{\rm tR} \ .
   	  \label{h}
   \end{equation}
The  first   term in (\ref{h}) is 
$
\hat H_{\rm ac} = \omega_0\hat a^\dagger\hat a +\epsilon  \hat a^\dagger\hat a(\hat a^\dagger\hat a-1) $; it  governs cavity field dynamics.
  $\hat a^\dagger$ and $\hat a$ are, respectively, boson creation and annihilation operators   acting in a basis of Fock states   $|n\rangle$ with different photon numbers, and the nonlinearity $\epsilon$  defines   Bose-Hubbard  interaction  strength. 
 $\hat H_{\rm L}=\sum_k E_{{\rm L},k} \hat b^\dagger_k \hat b_k $ and $\hat H_{\rm R}=\sum_ p E_{{\rm R},p} \hat c^\dagger_p \hat c_p$  describe  photon dynamics in the  left and right waveguides, respectively; $\hat b^\dagger_k$, $\hat b_k $ and  $\hat c^\dagger_p$, $\hat c_p $ are  creation and annihilation operators acting in spaces of modes labeled by a quasimomentum $ k  $ and $ p $ in a respective waveguide.  Couplings   between the cavity and   waveguide    modes    are assumed   weak, hence, the respective Hamiltonians are postulated in 
the form corresponding to the rotating wave approximation,
   $\hat H_{\rm tL}=t_{\rm L}   \sum_k    \hat b_k^ \dagger \hat a + ({\rm H.c.}) $ and $ \hat H_{\rm tR}=t_{\rm R}  \sum_p    \hat c_p  ^ \dagger \hat a+ ({\rm H.c.}) $,
   where $t_{\rm L}$ and $t_{\rm R}$ are coupling  amplitudes. Broad spectra of  $E_{{\rm L},k}$ and $E_{{\rm R},p}$ determine   densities of states, $\nu_{\rm L}$ and $\nu_{\rm R}$,  at the cavity mode $  \omega_0$ and   relaxation rates, $\Gamma_{\rm L}=\pi \nu_{\rm L} |t_{\rm L}|^2$ and $\Gamma_{\rm R}=\pi \nu_{\rm R} |t_{\rm R}|^2$.

    The measured photon current  is defined as the  average number of photons  $n_{t_0}$  that have passed during the measurement time 
         $t_0$, $J=\frac{1}{t_0}\langle \hat n_{t_0}\rangle$.  Brackets denote an average for a quantum mechanical operator $\hat{\mathcal{O}}$  with a nonequilibrium density matrix $\hat\rho$, $\langle{\hat{\mathcal{O}}}\rangle={\rm Tr}[\hat{\mathcal{O}}\hat\rho]$. The trace is calculated with the help of the nonequilibrium Keldysh  technique and  path integrals over 
        complex boson fields.      For the  limit (\ref{cond}), the   wideband drive induces the current  $J=2F\frac{\Gamma_{\rm L}\Gamma_{\rm R}}{\Gamma}$ that  does not depend on the interaction   $\epsilon$.
  The average photon number in the cavity, $\langle \hat a^\dagger\hat a\rangle=F \frac{ \Gamma_{\rm L} }{ \Gamma} $, does not saturate as $F$ increases. This value of $\langle \hat a^\dagger\hat a\rangle$ might be associated with   effective temperature $T_{\rm eff}\sim F \frac{ \Gamma_{\rm L} }{ \Gamma}\omega_0$, however, the incident photons are not distributed by the Bose-Einstein law and  the cavity has non-Gibbs density matrix.
 
 Zero frequency noise of the photon current is defined as the second cumulant of counted photons during   $t_0$ interval, $S=\frac{1}{t_0}\left(\langle n_{t_0}^2\rangle-\langle n_{t_0}\rangle^2\right)$. 
 Below we show that the noise-current relation, $S(J)$,    and    transmission probability function $T_\omega$, which appears in a photonic counterpart of the Landauer formula,  depend nontrivially  on $\epsilon$.  These quantities, as well as $g^{(2)}$-correlator and Fano factor, demonstrate  the richness of  Lorentzian-to-Gaussian crossover in a pseudothermal light.

 \section*{Results} \subsubsection*{Non-equilibrium effective action}
  \begin{wrapfigure}{r}{0.5\linewidth}
		\vspace{-0pt}\centering 
		\includegraphics[width=12cm,height=6cm,keepaspectratio,]
		{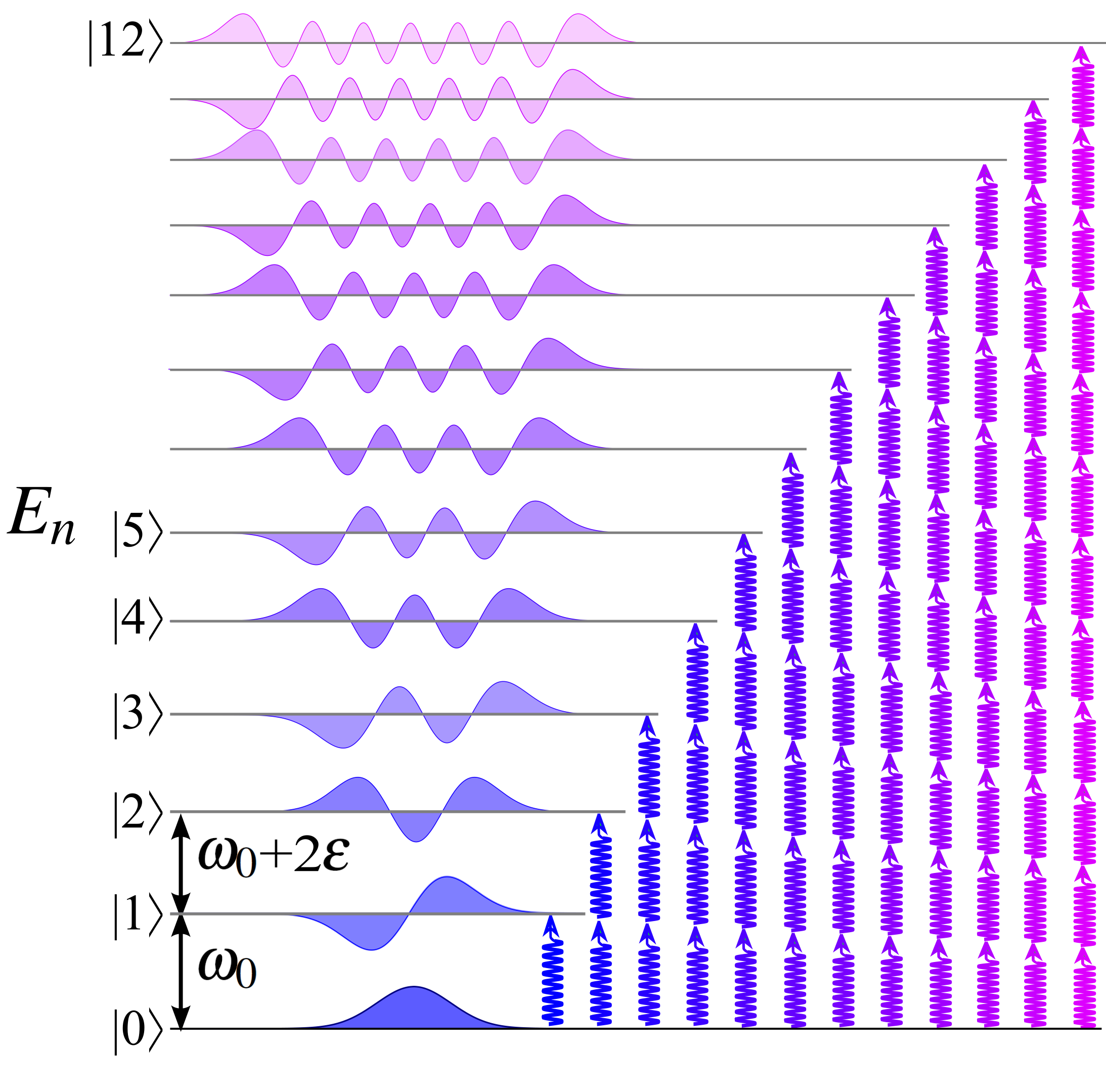} 
	\caption{
		Sketch of multi-photon transitions in the cavity with a negative anharmonicity $\varepsilon<0$.   Eigenstates $| n\rangle$ have non-equidistant spectrum $E_n=\omega_0 n +\epsilon n(n-1)$  where $n$ is the respective photon number. A multi-photon transition from $|0\rangle$ to $|n\rangle$ is accompanied by the absorption of $n$ photons of  the frequency $\omega_n=\omega_0+\epsilon (n-1)$. The decrease  of $\omega_n$  with $n$  for  $\varepsilon<0$ is illustrated by a change of color from blue to magenta.\vspace{0pt}
}  \label{multi-photon} \end{wrapfigure}
We reduce the  description to the Keldysh effective action $\mathbf{S}_{\rm eff}[\phi]$ for  
  the Hubbard-Stratonovich  real field $\phi(t)$. It  decouples the Bose-Hubbard interaction term. The dynamics of $\phi(t)$ determines the  decoherence of transmitted photons.    A particular  configuration of   $\phi(t)$ has a contribution $\sim e^{i\mathbf{S}_{\rm eff}[\phi]}$ in   the partition function. 
 The dynamics of  $\phi(t)$ is influenced by  different   multi-photon transitions in the cavity  shown  in Fig.~\ref{multi-photon}. The particle number  operator $\hat n=\hat a^\dagger \hat a$ does commute with the cavity Hamiltonian, $[\hat H_{ac},\hat n ]=0$. Hence, its Fock states $|n\rangle$ are the same  as those of the harmonic oscillator,   {\it i.e.}, they can be represented through the canonical coordinate and  Hermite polynomials. The wavefunctions in canonical coordinates  should not be confused with  multi-mode field configurations in the real space. Contrary to 
 harmonic oscillator, the eigenvalues spectrum of  $\hat H_{\rm ac}$ is a non-equidistant, $E_n=\omega_0 n+\epsilon n(n-1)$.
   It is shown that a description of  multi-photon transitions in the nonlinear single-mode cavity is reduced to an effective theory 
    for $\phi(t)$. It has a stochastic behavior, which resembles  
   Kubo model\cite{doi:10.1002/9780470143605.ch6} of an oscillator with random modulated frequency.
     For the sake of compactness, the results of this Section are presented  for the symmetric setup with $\Gamma_{\rm L}=\Gamma_{\rm R}=\Gamma/2$.  
   We find that  the stationary part of $\phi(t)$   is a nonequilibrium  saddle point of the  action. Its value defines   a shift of  the cavity mode frequency    by the occupation number $F$
     and the anharmonicity,  $
   	\omega_{\rm ac}= \omega_0+\frac{1}{2}(F-2)\epsilon  
	$. 
	A nonperturbative solution for the decoherence follows  from     Gaussian theory for 
	fluctuations  near the saddle point. In our  approach, 
	we distinguish the fields $\phi_+(t )$ and $\phi_-(t)$ residing on  the upward and backward   branches of the Keldysh contour $\mathcal{K}$.  
	 Hence, the stochastic  and quantum  phases are introduced as
	$\Phi(t) =\frac{1}{2}\int^t ( \phi_+ (t') + \phi_-(t') )dt' $ and $ \varphi(t) =\frac{1}{2}\int^t ( \phi_+ (t')  - \phi_-(t') )dt'$, respectively.
Their dynamics 
is 
governed by the   action of Caldeira-Leggett type\cite{Caldeira:1981aa,Shnirman:2016aa}, 
\begin{equation}
	i\mathbf{S}_{\rm CL}[
	\Phi,\varphi] =  
	\int \frac{i\omega^2}{\varepsilon} \Phi_{-\omega}  \varphi_{\omega}\frac{d\omega}{2\pi} \  +  \  
	\frac{i}{2}\int\frac{\omega^2d\omega}{2\pi} 
	\begin{bmatrix} \Phi_{-\omega} & \varphi_{-\omega}
	\end{bmatrix}
	\begin{bmatrix} 0&& \alpha^A_\omega \\ \\
		\alpha^R_\omega && i\alpha^K_\omega 
	\end{bmatrix}
	\begin{bmatrix}  \Phi_{\omega} \\ \\  \varphi_{\omega} 
	\end{bmatrix}   \ , \label{S-eff-1}
\end{equation}
where Fourier transformed   phases read as $\Phi_{\omega}=\int \Phi(t)e^{i\omega t} dt$ and $\varphi_{\omega}=\int \varphi(t)e^{i\omega t} dt$.
 The first term is  the 'kinetic' 
 part, whereas the    second term is Caldeira-Leggett      dissipative part.  It is  written as a matrix with Keldysh causality structure.

  In the equilibrium situation,  the many-body density matrix is $\hat \rho=\frac{1}{{\rm Tr} [e^{-i \beta \hat H}]}e^{-i \beta \hat H}$, where $ \beta$ is  the  inverted temperature. The fluctuation-dissipation theorem holds in this case.
It states that 
 the Keldysh component $\alpha^K_\omega$ and  retarded (advanced) components  $\alpha^{R(A)}_\omega$ in the effective action (\ref{S-eff-1}) are related as   $\alpha^K_\omega=(\alpha^R_\omega-\alpha^A_\omega)(1+2N_{\rm B,\omega})$, where $N_{\rm B,\omega}=\frac{1}{2}\coth \frac{\beta\omega}{2}-\frac{1}{2}$ is   bosonic occupation number.  In our  situation with flat 
   distribution functions, we find that dissipative   terms 
    vanish in the   limit of large $ \Delta$,  $\alpha^R_\omega=(\alpha^A_\omega)^*=\frac{  4i}{\pi \Delta^3} \epsilon \Gamma F^2   $. Oppositely, the fluctuational Keldysh term does not vanish and depends non-linearly on the occupation number  $ F$, $\alpha^K_\omega =  
   4 \frac{ \Gamma F (F+2)}{   4 \Gamma ^2+\omega ^2 }$. 
The distribution function in the anharmonic cavity is found, $N_{\rm ac,\omega}=\frac{1}{\Gamma}(\Gamma_{\rm L}N_{\rm L,\omega}+\Gamma_{\rm R}N_{\rm R,\omega})$. The inequality $\alpha^K_\omega \neq (\alpha^R_\omega - \alpha^A_\omega)(1+2N_{\rm ac,\omega})$ demonstrates a break of the fluctuation-dissipation theorem  in   our nonequilibrium  regime.

  \subsubsection*{Decoherence}
 Caldeira-Leggett action 	is reduced  to a Langevin equation\cite{Eckern:1984aa}   $\frac{d}{dt} \Phi(t)= \epsilon \xi(t)$ for the stochastic phase of transmitted  photons. The random force in the r.h.s. has the following correlator,   $\langle\xi(t)\xi(0)\rangle=\frac{1}{4}\alpha^K(t)$, where 
 $\alpha^K (t)=   F(F+2) e^{-2 \Gamma  |t|}$  is the time-resolved Keldysh part from (\ref{S-eff-1}). 
 The non-Gaussian effects of the noise 
    can be sensitive to the dynamics of $\varphi$; this is beyond the scope of our consideration. 
   The non-Gaussian effects might be important in the limit of large $\epsilon>\omega_0$ when the anharmonic cavity becomes a two-level system.

The decoherence of photons is determined  by the envelope $z(t)=\langle e^{i\Phi(0)-i\Phi(t)}\rangle=e^{-D(t)}  $ where $D(t)=\frac{1}{4}\int^t_0\int^{t'}_0\alpha^K(t'')dt''dt'$ is the  phase autocorrelation function. It reads  
\begin{equation} 
	z(t)  = {\rm exp}\Big[- 
	\frac{\kappa^2}{2\Gamma^2}(2\Gamma t +  e^{-2 \Gamma  |t|}-1  )\Big]   \ .
	\end{equation}
The same structure of 
 a correlator appears in the   Kubo model. 
 There is  Gaussian decay at short timescale with the  rate  
$\kappa = \frac{1}{2}\epsilon\sqrt{F(F+2)}$,  which
increases with the anharmonicity and  is nonanalytic by $F$ due to the square root. 

The behavior of $z(t)$ plays a central role in our findings because it determines all characteristics of the outgoing  light.   We note that the Gaussian decay of the envelope  appears  in the spin-boson model with  the sub-Ohmic dissipative environment
   with $1/\omega$ spectrum~\cite{Shnirman_2002}. The effective sub-Ohmic spectrum of phase fluctuations  
 in our problem is  due to    multi-photon transitions.

\subsubsection*{Intensity correlator}

   The second-order correlator    of outgoing photons,  which can be probed in Hanbury-Brown and Twiss interferometry\cite{Gabelli:2004}, is    
   \begin{equation}
   g^{(2)}(t)=1+z^2(t)e^{-2\Gamma t} \ .
   \end{equation}  Thus, the emitted signal is chaotic pseudothermal light  that can be 
   Lorentzian  or Gaussian depending on   the anharmonicity  $\epsilon$ 
   and  
  mean photon  current. 
  Namely, the Lorentzian light with $g^{(2)}(t)\approx 1+e^{-2 \Gamma  t}$ occurs at $J\ll J^*_\epsilon$,   whereas the  Gaussian,  $g^{(2)}(t)=1+e^{-2 \kappa^2  t^2}$,  at $J\gg J^*_\epsilon$  is found. The value of $J^*_\epsilon$  is 
   related to  the ratio between the anharmonicity and relaxation as $J^*_\epsilon=\frac{\Gamma}{2}\Big(\sqrt{1+4\frac{\Gamma^2}{\epsilon^2}}-1 \Big)$.

  The correlator at zero time $g^{(2)}(0)>1$ is related to the  bunching and $g^{(2)}(0)<1$ to the antibunching of emitted photons\cite{Zou:1990aa}. 
In our case, $g^{(2)}=2$ that is consistent with that pseudothermal photons are bunched.  This occurs because the incoherent drive with a flat spectrum resembles the high-temperature limit of Bose-Einstein distribution  where photon number fluctuations in a single mode are proportional to mean photon number squared.  

However, the presence of coherence in the emitted light suppresses positive correlations. As shown below this is reflected 
in the emergence of the shot noise and in 
the suppression of the Fano factor. This is also indicated by the fast decay of  $g^{(2)}(t)$ to the unity, which is associated with  a coherent light. The decay occurs at the timescale   $t\sim 1/\kappa$, which becomes very short at large $F$ and $\epsilon$.

  \subsubsection*{Transmission spectrum}
 
 It is found that the single-photon Breit-Wigner transmission function $ \tau_\omega= \frac{
	\Gamma^2}{(\omega - \omega_{\rm ac})^2+\Gamma^2 }$ (exactly-solvable  $\epsilon=0$ limit) is modified by the stochastic phase as  
  $	T_{\omega}=\int\limits^{\infty}_{-\infty}  	z(t) \tau(t) e^{i \omega  t } dt$.	  Here, $\tau(t)$ is time-resolved $\tau_\omega$. 
 Similarly to $g^{(2)}$, the  Lorentzian transmission spectrum $\tau_\omega$ is changed to Gaussian at $J\gg J^*_\epsilon$ which reads
 \begin{equation}
 	T_\omega=\frac{\sqrt{\pi}\Gamma  }{2  \kappa } \exp\left(\!-\frac{(\omega-\omega_{\rm ac})^2}{4\kappa^2 } \right)  \ .  \end{equation}This is a nonperturbative result with respect to the interaction and incoherent drive.   This transmission describes  nonequilibrium  and  
many-body    
  photon transfer.

\begin{wrapfigure}{r}{0.5\linewidth}
		\vspace{0pt}\centering 
\includegraphics[width=12cm,height=7.5cm,keepaspectratio,]{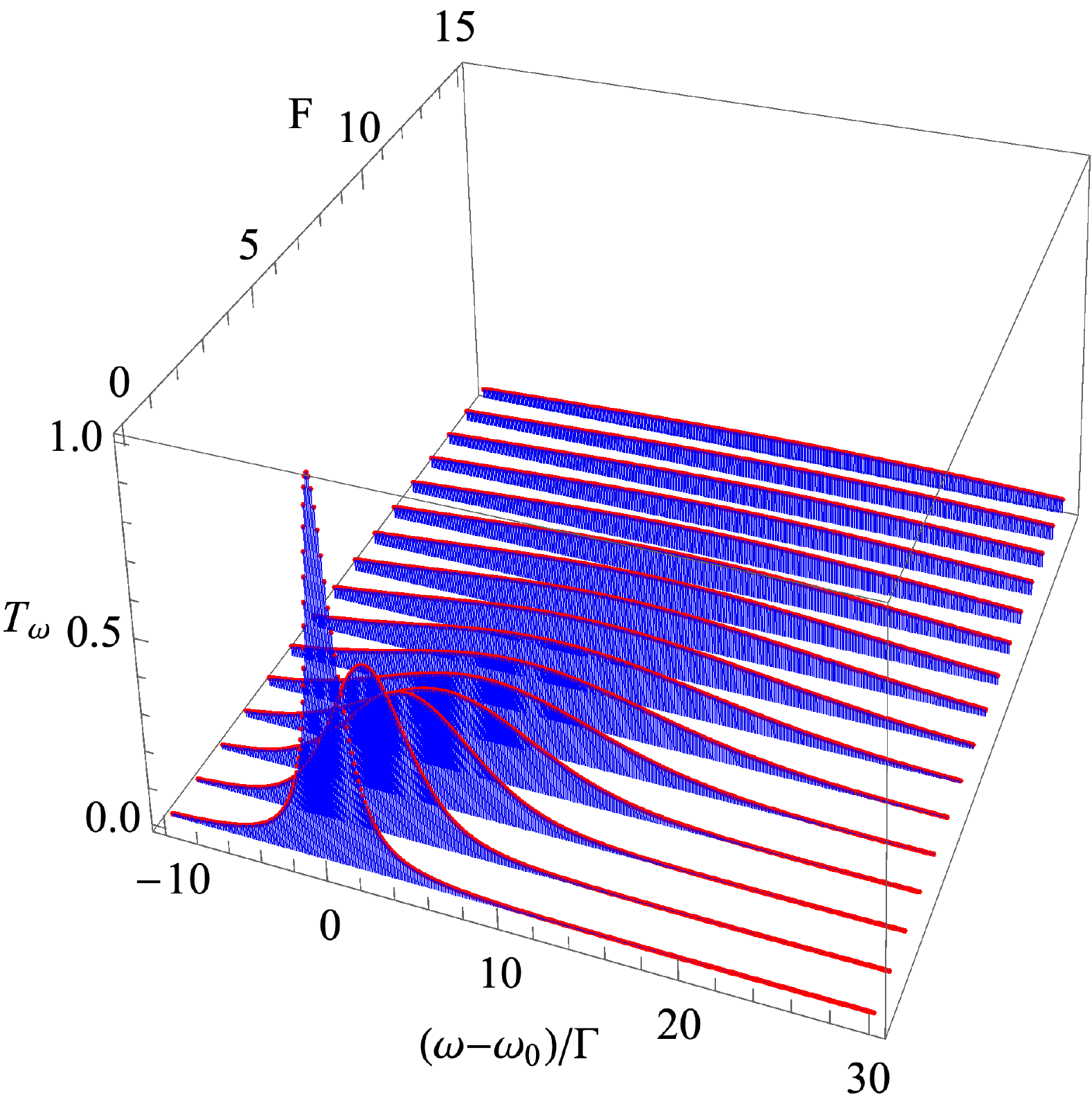}
	\caption{
Transmission spectrum $T_\omega$ as a function of the frequency  and   $F$.   The curve at equilibrium regime $F=0$ represents Lorentzian $\tau_\omega$  of the width $\Gamma$.
		The increase of $F$ turns the nonequilibrium regime on and $T_\omega$   become Gaussian with $\kappa> \Gamma$.  Maxima of  $T_\omega$ at each $F$ are located at shifted frequency  $ F\epsilon/2
		$; here, $\Gamma_{\rm L}=\Gamma_{\rm R} =\Gamma/2$ and 
		   $\epsilon/\Gamma=2$.
} \vspace{-20pt} \label{transmission}\end{wrapfigure}

The crossover between Lorentzian and Gaussian spectra in $T_\omega$, when $F$ increases, is shown in Fig.~\ref{transmission}; results are obtained after numerical integration. It should be noted that the photon-photon interaction redistributes the spectral weight of $T_\omega$ around  $\omega_{\rm ac}$,  keeping  its integral value constant. The latter explains that  fact that the anharmonicity does not change the current in the wideband input drive regime.

\subsubsection*{Noise  
} 
    Correlations in the output field are determined by  $g^{(2)}(t)$. A relevant  information is given by a short timescale of $ t_0\ll 1/\Gamma$. 
 Distinctly, the zero-frequency noise\cite{BLANTER20001}  contains information about the long   timescale, $t_0\gg 1/\Gamma$.
 The indicated change of  the asymptotic behaviors  in $z(t)$ or $T_\omega$ 
 is reflected   in the richness of a low-frequency noise-current relation $S(J)$. 

\begin{figure}[h!]
	\centering
	\subfigure[Noise-current scaling\label{diagramnoise}]{\includegraphics[width=15cm,height=6.85cm,keepaspectratio,]{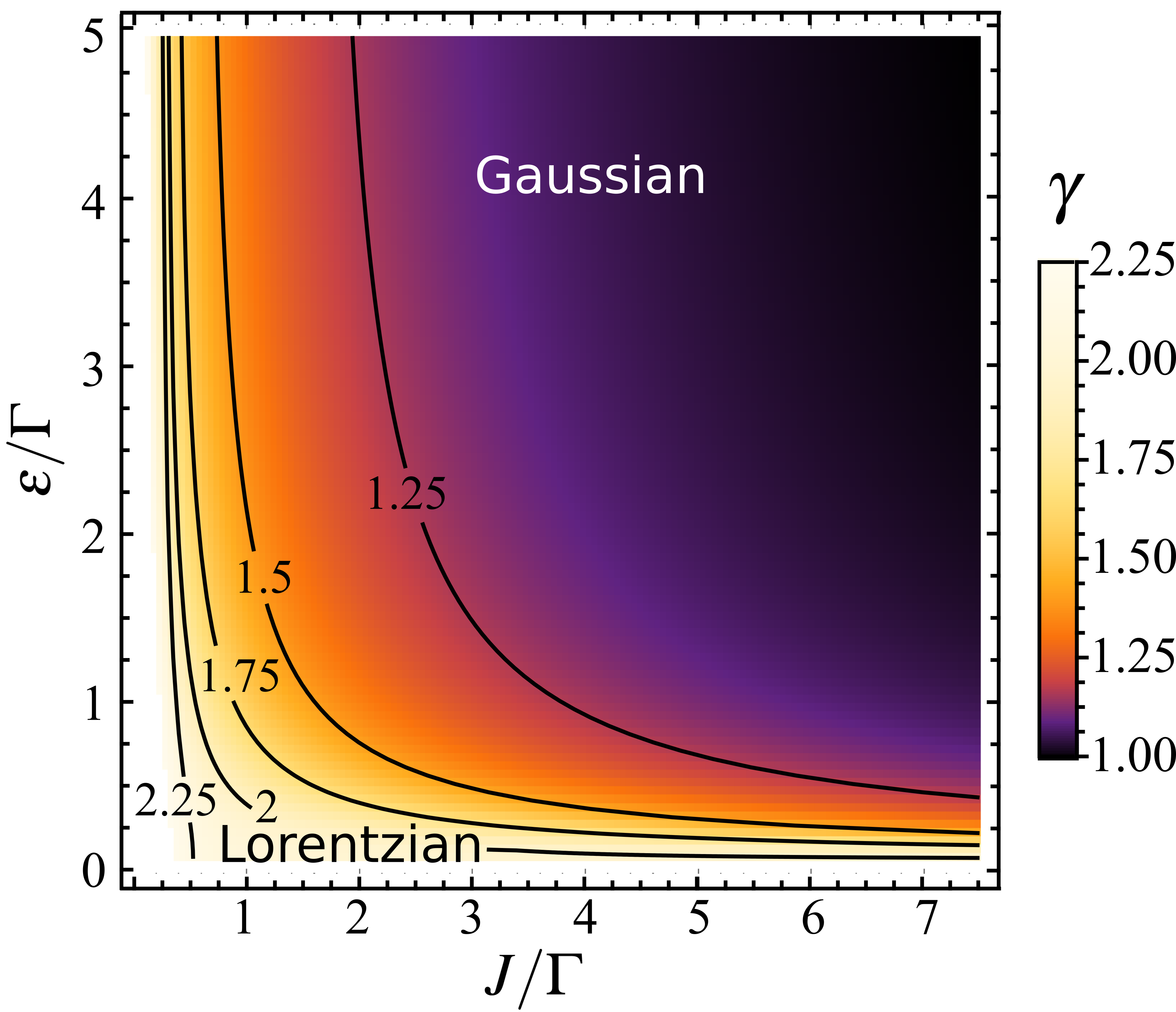}} \quad \quad 
\subfigure[Fano factor\label{fanofactor}]{\includegraphics[width=13cm,height=7cm,keepaspectratio,]{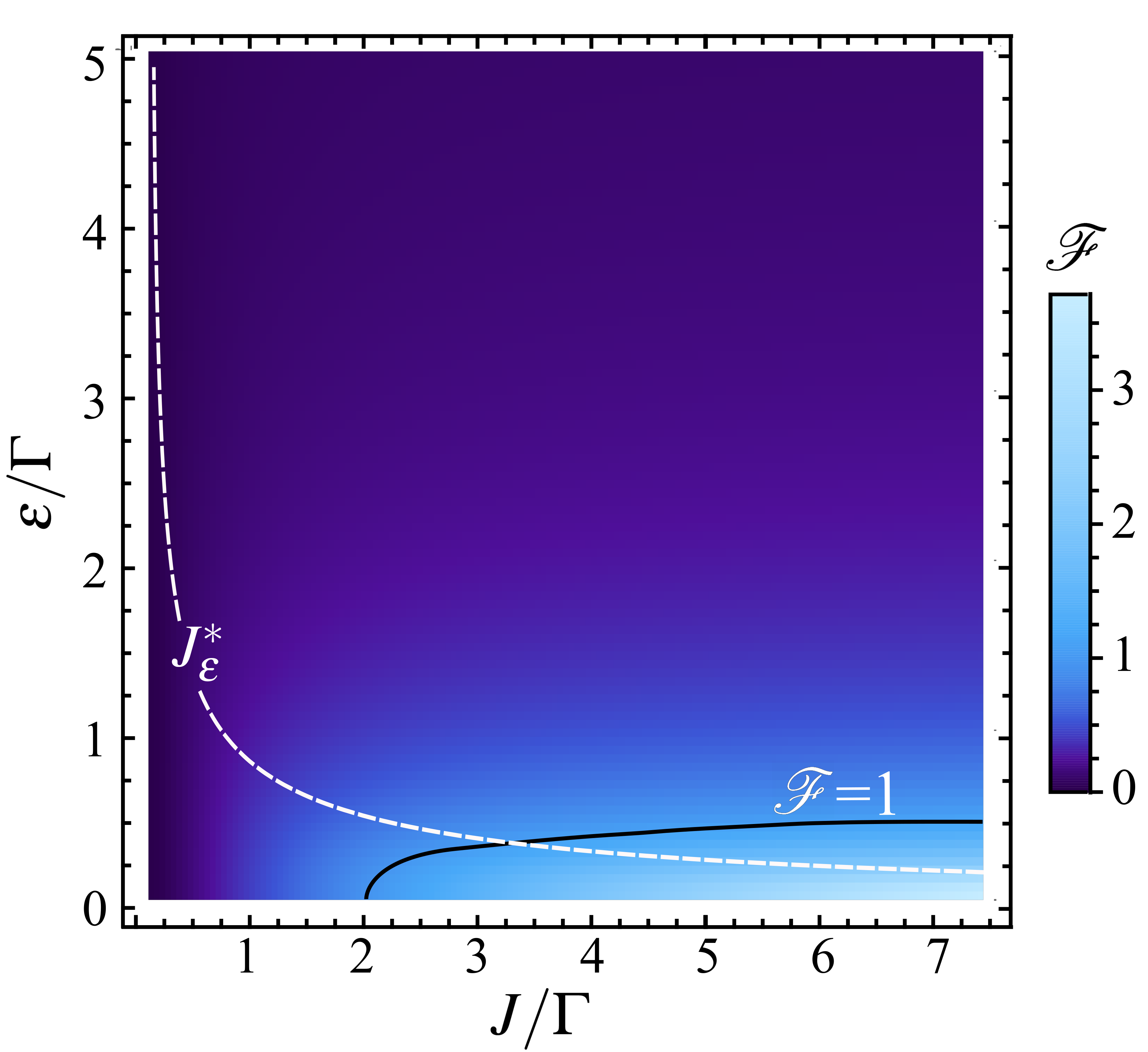}} 
	\caption{(a) Results 	for  scaling exponent $\gamma$  in noise-current relation $S\propto J^\gamma$ as   functions of  the photon current $J$ and anharmonicity $\epsilon$.  Color map  represents logarithmic derivative $\gamma=\frac{d \ln S}{d\ln J}$.    
	Contours   of constant $\gamma$ show a crossover between 
		Lorentzian and Gaussian pseudothermal light. The Lorentzian light corresponds to single-particle transport with Breit-Wigner   transmission and $\gamma=2$,  and the Gaussian  is to many-body transport and the shot noise. 
		The   fractional  $\gamma=3/2$ asymptotic appears at low currents 
		and large anharmonicity $\epsilon/\Gamma>1$.  
(b)~Color map for Fano factor $\mathcal{F}= S/J$. The white dashed curve is $J^*_\epsilon$. The black solid line stands for $\mathcal{F}=1$ contour.
} 
\end{figure}

 The Lorentzian emitted light shows the quadratic noise-current relation, 
 \begin{equation}S_{\rm therm} 
 =\frac{J^2}{2\Gamma} \ , \quad J\ll J^*_\epsilon \ .
 \end{equation} 
  This scaling  resembles a situation of equilibrium radiation in a  cavity  at  a very high temperature  where  the  variation, $\langle\!\!\langle n^2\rangle\!\!\rangle=\langle n^2\rangle - \langle n\rangle^2$, and average  photon number  are related as  $\langle\!\!\langle n^2\rangle\!\!\rangle \sim \langle n\rangle^2$.
The  upper boundary for the current $J^*_\epsilon$ can be large or small compared to $\Gamma$ depending on the ratio $\epsilon/\Gamma$. 
 As shown in Fig.~\ref{diagramnoise}, the  region of small   $\varepsilon$  corresponds to a  Lorentzian light with the   exponent $\gamma \approx 2$ in the noise-current relation  represented as $S\propto J^\gamma$.
We note that the quadratic dependence  and  the Lorentzian   spectrum of $S_{\rm therm} $ at finite $\omega$ (see Section "Generating functional method" and Eq. (\ref{S-0}) for details), is   analogous to    a modulation noise    in quantum point contacts\cite{Kogan:2008aa}.

In the many-body regime with Gaussian light the following expression  is found 
\begin{equation}
S=\frac{\sqrt{\pi}}{2\sqrt2}\frac{J\Gamma}{\epsilon\sqrt{1+\Gamma/J}}
\ , \quad J\gg J^*_\epsilon \ . \label{S-main}
\end{equation}
The absence of the quadratic scaling in $S(J)$ indicates the shot noise behavior.  Also, this means that the Gaussian pseudothermal light has a partial coherence. 
In the limit of strong anharmonicity, $\epsilon\gg \Gamma$, the scale $J^*_\epsilon\sim\frac{\Gamma^3}{\epsilon^2}$ is small compared to $\Gamma$. Consequently,  the ratio $\Gamma/J$ in the square root of (\ref{S-main}) can be both large or small compared to unity. This shows that two different asymptotical regimes of the Gaussian noise do  exist, they are associated with $\Gamma/J\to 0$ and $\Gamma/J\to \infty$ limits. These limits have a continuous crossover between each other at $\Gamma/J\sim 1$.

If the  drive is sufficiently strong, such that   $J\gg\Gamma$ holds, then   
we arrive at the  linear shot noise-current relation
$S_{\rm shot} =\frac{\sqrt{\pi}}{2\sqrt2}\frac{\Gamma}{\epsilon}J$.
We note that    the scalings  similar to   $S_{\rm shot}\propto  J$ and  $S_{\rm therm}\propto  J^2$     were discussed in Refs.~\cite{Beenakker:1999aa,Beenakker:2000aa}. Namely, the  photon current  emitted by a  coherent source and propagated through a random media     shows a linear  noise-current relation  if the scattering region is absorbing, and quadratic if it is amplifying.

At low currents,  $\Gamma\gg J \gg J^*_\epsilon
$, the ratio  $\Gamma/J$ in (\ref{S-main}) dominates over the unity  and we arrive at      unconventional  
scaling of the noise, 
$S_{\rm shot}'\propto J^{3/2}$. This  is one of the remarkable findings of this work. The change of the scaling exponent in   $S\propto J^\gamma$  from $\gamma=2$ to $\gamma=3/2$  as $\epsilon$ increases  can be understood as emerging of a partial coherence in the propagated light.  The difference between these nonequilibrium fixed points   is illustrated by blue and   orange curves in Figs.~\ref{SJ} and \ref{exponent}. 
In the limit of weak anharmonicity, $\epsilon\ll\Gamma$, the asymptotic behavior   of $S\propto J^{3/2}$ 
does not appear. Here, we obtain a simple crossover between the thermal-like behavior and the shot noise. It is demonstrated as blue curves,  which start at $\gamma=2$ and  saturate  smoothly at $\gamma=1$. 
Parameter domains     corresponding to different  asymptotic  limits  are collected in Table~\ref{tab1}.

\subsubsection*{    Fano factor} 
It is instructive to analyze the Fano factor,  the ratio of the low-frequency   noise  and current,   $\mathcal{F}=S/J$. 
In   mesoscopic physics,   $\mathcal{F} $ defines an effective charge that is transmitted  coherently. This applies, e.g., to  single-electron transistors\cite{Choi:2003aa}, to helical electrons scattering  \cite{Nagaev:2018aa,Kurilovich:2019aa}  and   to  multiple Andreev reflection in superconducting junctions\cite{Cron:2001aa}. In our optical system, it can be  associated with the  emergence of positive ($\mathcal{F} >1$) and negative  ($\mathcal{F} <1$) correlations between  photons\cite{Shapiro:2020aa,Shapiro:2019aa}.

In the regime of conventional shot noise we find universal result $\mathcal{F}_{\rm shot}=\sqrt{\frac{\pi}{8}} \frac{\Gamma }{ \epsilon} \ll 1$ that does not depend on the drive amplitude.
The unconventional   shot noise $S_{\rm shot}'\propto J^{3/2}$ corresponds to Fano factor $\mathcal{F}'_{\rm shot}=\sqrt{\frac{\pi}{8} \frac{\Gamma J}{ \epsilon^2}}$ which is also  smaller than unity. 
 The regime of Lorentzian light  yields $\mathcal{F}_{\rm therm}= \frac{J}{2\Gamma} $ which can be less or greater than unity.
  Thus, the Fano factor represents   the complexity of   nonequilibrium properties  of the emitted light.    Its color plot, the  contours with $\mathcal{F}=1$ (black curve), and $J^*_\epsilon$ (white dashed curve) are presented in Fig.~\ref{fanofactor}.

\section*{Discussion and outlook
}\label{Discussion}

Studies of driven-dissipative   optical  systems with Bose-Hubbard interaction is a large research area. 
In particular,    incoherent drive effects were  explored in experiments on the photon blockade~\cite{PhysRevLett.107.053602,Photon_Blockade_corr} and  thermal states propagation~\cite{Goetz:2017aa}. The photon blockade   is characterized by the antibunching of photons and their sub-Poissonian  statistics. The thermal states propagation is an alternative regime, which is characterized by a suppressed photon blockade. In that experiment~\cite{Goetz:2017aa},   the    bunching effect and super-Poissonian  statistics of emitted photons were observed. In this  work, we investigated how the incoherent pump influences   the photon transport   in a wideband limit and arbitrary excitations number. Namely, the central question of this work is how does a photonic counterpart of thermal transport through an electronic quantum dot with a non-equidistant spectrum  look like? This question seems actual in  view of increasing interest to novel phenomena and phases in     nonequilibrium cavity and circuit QED  where the fluctuation-dissipation relation and detailed balance condition are violated.

\begin{wrapfigure}{r}{0.5\textwidth}
   \centering	{\subfigure[Noise-current relation \label{SJ}]{\hspace{-4pt}	\includegraphics[width=1\textwidth,height=5.2cm,keepaspectratio,]{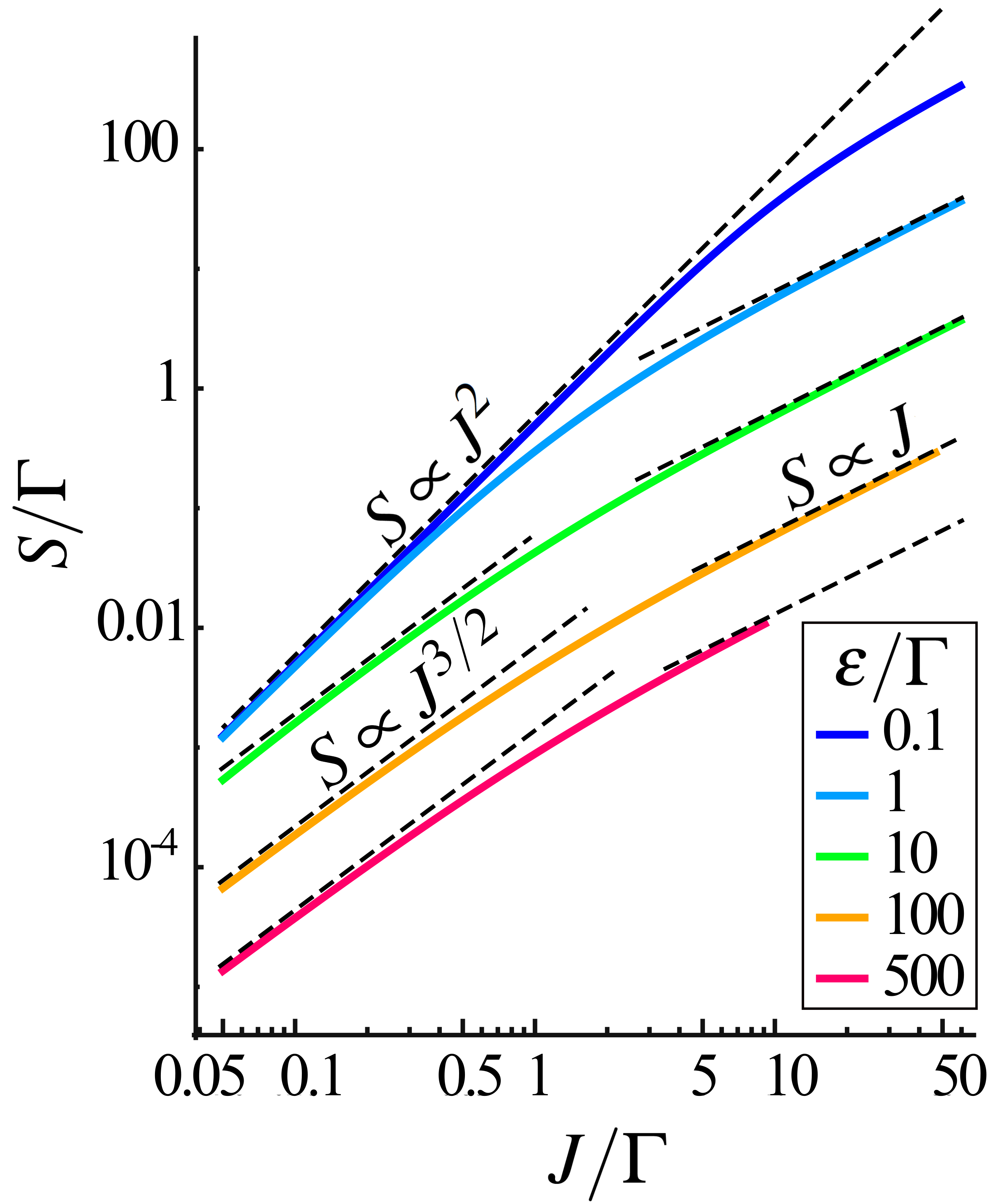}}  
   \subfigure[Scaling exponent \label{exponent}]{\hspace{5pt}\includegraphics[width=1\textwidth,height=5.2cm,keepaspectratio,]{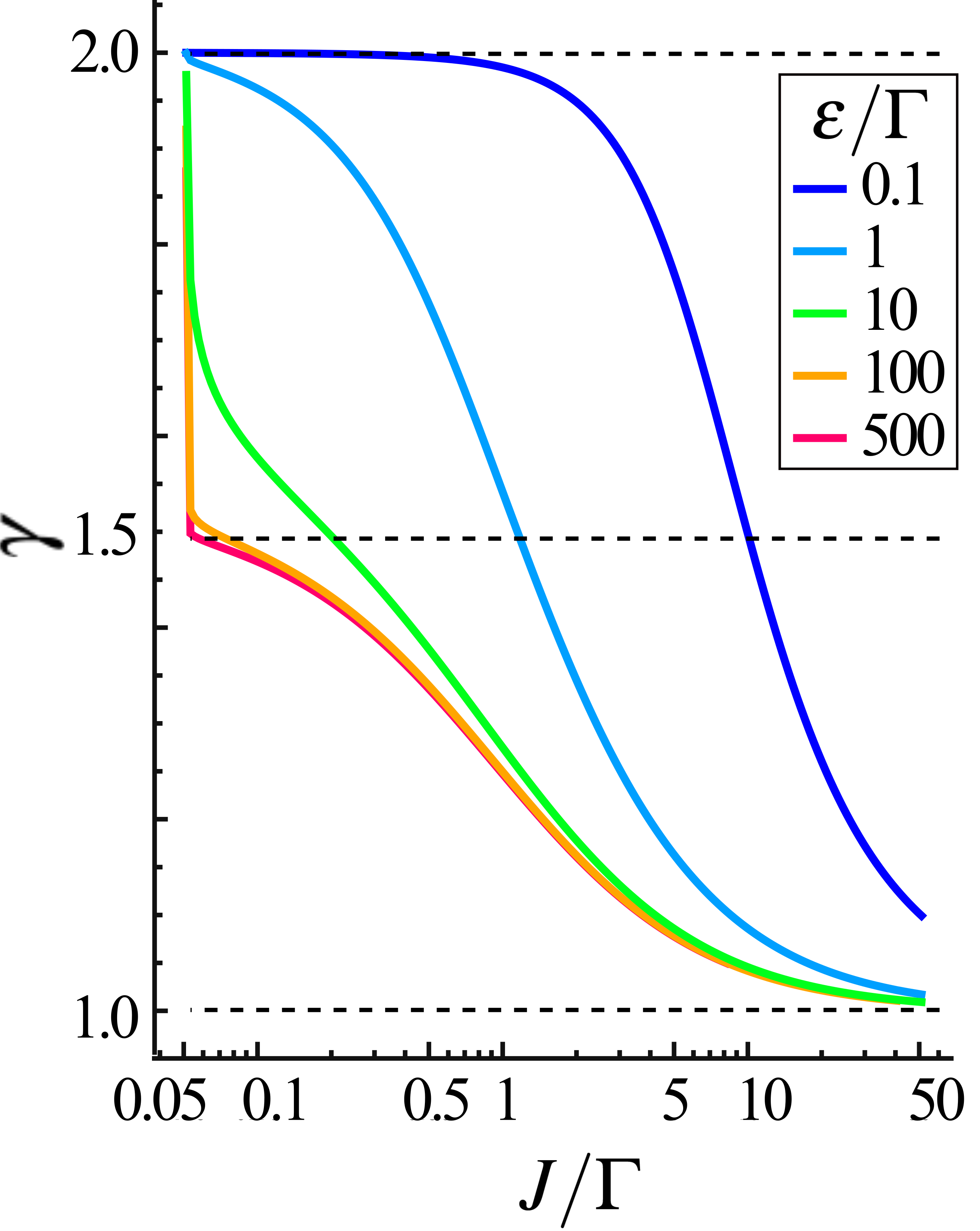}}}
   	\caption{(a)~Noise-current relation $S\propto J^\gamma$ for weak ($\epsilon/\Gamma=$0.1), intermediate ($\epsilon/\Gamma=$1) and strong  ($\epsilon/\Gamma=$10, 100, 500) anharmonicity (Bose-Hubbard interaction strength).
	 The logarithmic scale demonstrates three different regimes of the photon transport: quadratic thermal behavior at weak interaction, and fractional  power-law $S\propto J^{3/2}$ at small currents that changes to conventional shot noise $S\propto J$ as $J$ increases.   (b)~Scaling exponent $\gamma$ that demonstrates a change of fixed point from $\gamma=2$  to $\gamma=3/2$ as the  interaction increases.
   } \vspace{-20pt}  \end{wrapfigure}

The bandwidth of the  drive signal is  assumed to be greater than  characteristic relaxation rates. In this limit, we  found that the  increase
of the Bose-Hubbard interaction and   incoherent pump induce an intriguing transition from effectively thermal fluctuations  to partially coherent nonequilibrium noise. 
A complexity  of this transition is demonstrated in Fig.~\ref{exponent} where  cross-sections of the phase diagram from Fig.~\ref{diagramnoise} are shown for particular values of the interaction.

First,  the Lorentzian  transmission function and noise spectrum are transformed into Gaussian. 
The latter is a signature of   many-body interaction in the anharmonic cavity    and the consequence of   transmitted photons decoherence.  The time-resolved intensity correlator $g^{(2)}$  and   photons dephasing  reveal transitions from exponential to Gaussian decay laws demonstrating a certain similarity with Kubo model of an oscillator with stochastic frequency and also with the spin-boson model with sub-Ohmic $1/\omega$ spectrum.

	Second, the low-frequency noise $S$ and  the Fano factor are nonlinear functions of the drive amplitude, or, the average transmitted current. The scaling exponent  
	in the noise-current ratio, $S\propto J^{\gamma}$, can have three universal values. They indicate  (i)  single-photon transfer  and effectively thermal state noise with $\gamma=2$, and (ii) partial coherence in strongly nonequilibrium limit with the shot noise ($\gamma=1$) and  suppression of the Fano factor that becomes smaller than unity,  and (iii)  unconventional noise-current relation with the   fractional  exponent $\gamma=3/2$. 
	In the regime (iii), that is found for low currents,  the fluctuations of outgoing states are contributed by two competing components, the  coherent emission  and pseudothermal dissipative  dynamics   in a large Fock space of the cavity mode.  This regime is supposed to be relevant for experiments reported  in the Ref.\cite{Goetz:2017aa}, where the difference between fluctuations in thermal   radiation and  the Poissonian regime 
	was observed, and also in Refs.\cite{Dmitriev:2019aa,Honigl-Decrinis:2020aa,Zhou:2020aa}, where the statistics of incident photons was measured by a continuous wave mixing.
	
	In our solution, we found a single maximum in the transmission spectrum at the frequency  $\omega_{\rm ac}$. This is in   contrast to the  case of the coherent drive at a particular frequency, where      multi-photon  supersplittings\cite{Elliott:2016aa} are clearly observed in experiments\cite{braumuller2017analog}. 
	Their absence  in our solution is explained by   the averaging over a large amount of multi-photon transitions induced by the wideband and noisy drive.

The effective Caldeira-Leggett action proposed  here describes fluctuations in the Gaussian (quadratic) order approximation.  This approach provides a quasi-classical theory for noise  represented as symmetrized correlators  due to  the Keldysh time ordering.
	The non-Gaussian   fluctuations effects are the intriguing direction of recent studies related to non-linear quantum environment~\cite{PhysRevResearch.2.033196} and dynamical Coulomb blockade~\cite{Zhan_2020}; in particular, the possibility of different time orderings on  the Keldysh contour  becomes important in  the description of  quantum   noise   and nonsymmetrized correlators.

	\section*{Summary}
	\label{summary}
	To conclude, we investigated steady-state  
	photon transport  in the anharmonic single-mode oscillator. It can be transmon qubit or nonlinear optical cavity incoherently driven and coupled to the input and output waveguides. 
	We   predict a rich behavior of the noise and intensity correlators in the nonequilibrium. The nonequilibrium conditions reveal intriguing effects due to the photon-photon interaction, they include pseudothermal
	  fluctuations with  partial coherence,     Gaussian transmission spectrum with effectively   sub-Ohmic dephasing,  and unconventional shot noise 
	with fractional power-law scaling.

	The assumption of the wideband noisy drive field  allows for a compact analytical solution.  The decoherence  due to  the photon-photon interaction is described within a nonperturbative approach reduced to Caldeira-Leggett effective action. In this many-body problem, a stochastic   phase of transmitted photons  depends on  various multi-photon transitions. 
	We expect that the nontrivial behavior of   the noise    might not have a direct analogy in condensed matter systems.

	The  methodology    is based on the Keldysh field theory.  The solution for a system far from equilibrium is obtained. 	Our results give an insight into photon transport  
	in the cavity and circuit  QED, where  the  interaction and nonequilibrium dynamics  appear on an equal footing. They can be generalized, in particular, for driven-dissipative Bose-Hubbard lattices\cite{Vicentini:2018aa,Biella:2015aa}.
	It is suggested that these findings 	are experimentally accessible by methods based on  photon counting~\cite{PhysRevLett.107.053602,Baumann}, homodyne detection of intensity correlations\cite{birnbaum2005photon,Photon_Blockade_corr,Goetz:2017aa}, dispersive readout and spectroscopy~\cite{macha2014implementation,braumuller2017analog}.

\begin{table*}[ht]
	\caption{Noise regimes classification.}
	\centering
	\begin{tabular}{ | p{0.13\linewidth}
			| p{0.255\linewidth}| p{0.255\linewidth}| p{0.255\linewidth}|}
		\hline
		\vspace{0.1cm} Interaction  \vspace{0.1cm} & 
		\vspace{0.05cm} \raggedright Thermal  noise, $S_{\rm therm}\propto J^2$.  Lorentzian light \vspace{0.1cm} &  \vspace{0.0cm}  \raggedright Fractional scaling,    $S_{\rm shot}'\propto J^{3/2}$. Gaussian light \vspace{0.1cm} &  \vspace{0.05cm}   Conventional scaling,  $S_{\rm shot}\propto J $. Gaussian light  \vspace{0.1cm}  \\
		\hline
		\vspace{0.3cm}weak, $\epsilon\ll\Gamma$ &
		\vspace{0.0cm}$$J\ll\frac{\Gamma^2}{\epsilon}$$ & 
		\vspace{0.4cm} \centering{---} &  \vspace{0.0cm}$$ J\gg\frac{\Gamma^2}{\epsilon}$$  \\
		\hline
		\vspace{0.3cm}strong, $\epsilon\gg\Gamma$ &
		\vspace{0.0cm}$$J\ll\frac{\Gamma^3}{\epsilon^2}$$ & 
		\vspace{0.0cm}$$\frac{\Gamma^3}{\epsilon^2}\ll J\ll \Gamma $$ &  \vspace{0.2cm}$$ J \gg  \Gamma $$ \\
		\hline
	\end{tabular}
	\label{tab1}
\end{table*}

   \section*{Methods}
   \label{Methodology}
  In this part of the paper, we present the methodology of calculations. 
   We start from  a formulation of  nonequilibrium field theory that corresponds to a microscopic Hamiltonian (\ref{h}). 
   It is based on the Keldysh Green functions approach  and Hubbard-Stratonovich transformation decoupling  the non-Gaussian interaction term. 
In particular, the  Landauer  formula for photon  current is derived within the Keldysh formalism. Then, the  Caldeira-Leggett  action   is formulated and applied to calculations of decoherence, transmission function, noise spectrum,  and $g^{(2)}$-correlator.

   \subsection*{Effective functional 
   }

   \subsubsection*{Keldysh approach}
    The nonequilibrium field theory is started from the partition function   given by the path integral, $Z=\int D[\bar \Psi, \Psi]\mathcal{T} \exp(i\mathbf{S}[\bar\Psi, \Psi])$, where $\mathbf{S}=\int_{\mathcal K}  (i\bar \Psi\partial_t\Psi - H)dt$ is the  action determined on Keldysh time contour  $\mathcal{K}$. Here, $\mathcal{T}$ stands for Keldysh time ordering along   $\mathcal{K}$.  The vector with all path integral complex variables is $\Psi=[b(t,k),a(t),c(t,k)]^T$, and $\bar  \Psi=[\bar b(t,k),\bar a(t),\bar c(t,k)]^T$ is its conjugate.
   Let us   consider first   the Keldysh  action of the anharmonic cavity
   \begin{equation} 
   	i \mathbf{S}_{\rm ac}=i\int_{\mathcal{K}} \Big(\bar a(t)(i\partial_t-(\omega_0-\epsilon))a(t)-  
	\epsilon\bar a(t) a(t) \bar a(t) a(t) \Big)dt \ . \label{S-ao}
   	\end{equation} 
   By the virtue of Hubbard-Stratonovich real field $\phi$ we decouple the interaction term in   (\ref{S-ao}) as follows
   \begin{equation} 
   	\exp\Big(-i\int_{\mathcal{K}} \epsilon \bar a(t) a(t) \bar a(t) a(t) dt \Big) =  
	\int \!\!  D [\phi] 	\exp\Big(\int_{\mathcal{K}} \frac{i }{4\epsilon}\phi^2(t) dt - i\! \int_{\mathcal{K}} \bar a(t) \phi(t)  a(t) dt \Big) \ .
   \end{equation} 
    The last term in the exponent corresponds to fluctuating potential $\phi(t)$ that randomly modulates   oscillator  frequency. As a result of this transformation, the cavity action becomes the following:
    \begin{equation} 
    i\mathbf{S}_{\rm ac}  \to   \int_{\mathcal{K}} \frac{i }{4\epsilon}\phi^2(t) dt -  
     i\! \int_{\mathcal{K}} \bar a(t)\big(i\partial_t-(\omega_0-\epsilon +\phi(t) )\big)a(t) dt   \ .
   \end{equation} 
   With the use of this representation, the total action is formulated through matrix Green functions. Here, a transformation to the usual time axis $t\in [-\infty, \infty]$   doubles the number of variables, $\Psi(t_{\mathcal{K}})\to [\Psi_+(t), \Psi_-(t)]$ and $\phi(t_{\mathcal{K}}) \to[\phi_+(t), \phi_-(t)] $, where these fields reside on  upward or backward parts of the Keldysh contour labeled by   $\pm$ indices. This is associated with Keldysh $\check \tau$-space parametrized by Pauli matrices $\check \tau_{x,y,z}$ and identity matrix $\check \tau_0$. 
   
   After the standard Keldysh rotation to classical and quantum components for complex fields,
  $
   	\Psi_{c,q}(t)=\frac{1}{\sqrt{2}}\big(\Psi_+(t) \pm\Psi_-(t)\big) 
   	 	$, 
    	   and for real fields,
	  $\phi_{c,q}(t)=\frac{1}{2}\big(\phi_+(t) \pm\phi_-(t)\big)$, 
    we arrive at the following representation of the total action:
        \begin{equation}
    	i\mathbf{S}=  \int \frac{i }{\epsilon}\phi_c(t)\phi_q(t) dt + 
	i  \int  \bar \Psi
	(t)
    	\begin{bmatrix}
    	\check	G_{\rm L}^{-1} && -t_{\rm L}\check\tau_x && 0 \\ \\ 
    	-t_{\rm L}^*\check\tau_x  && \check G_{\rm ac}^{-1}[\phi] && -t_{\rm R}^*\tau_x  \\ \\ 
    	0 && - t_{\rm R}\check \tau_x && \check G_{\rm R}^{-1} 
    	\end{bmatrix}
     	\Psi
    (t) dt \ , \quad \Psi(t)=\begin{bmatrix}
      \Psi_c(t) \\ \\ 
      \Psi_q(t)
    \end{bmatrix}  \ . \label{S-total}
    \end{equation}      
Each element of the matrix is a  block in  $\check \tau$-space that possesses causality structure.
Let us consider the  left waveguide's block: 
\begin{equation}
	 \check G_{\rm L}^{-1}
	 = 
\delta_{k,k'}\begin{bmatrix}
	0 && i\partial_t-E_{{\rm L},k}- i0 
	 \\ \\ 
	i\partial_t-E_{{\rm L},k}+i0  && 2i 0 f_{{\rm L},k} 
\end{bmatrix} \ .
  \end{equation} 
   The retarded (advanced) Green functions are $G^{R(A)}=1/(\omega-E_{{\rm L},k}\pm i0)$), they differ by the sign of the infinitesimal   frequency shift $i0$ along the imaginary axis. The    Keldysh component   $2i 0 f_{{\rm L},k}$   encodes a distribution function  $N_{{\rm L},k}$ of incident photons  as $f_{{\rm L},k}=2N_{{\rm L},k}+1$. The similar definition applies for $\check G_{\rm R}^{-1} $.  The local in time  inverted Green function of the cavity 
    \begin{equation} 
    	 \check G_{\rm ac}^{-1}[\phi] = (i\partial_t-\omega_0+\epsilon -\phi_c(t) )\check \tau_x -\phi_q(t)\check \tau_0 
    \label{Gao}
    	\end{equation} 
 is nonstationary because it involves   $\phi_c(t)$ and $\phi_q(t)$.
   The nondiagonal elements $t_{\rm L,R}\check \tau_x$ in (\ref{S-total})   mix the cavity   and       waveguides' fields corresponding to $k$ and $p$ modes. 
   After the integration over   $b_k$ and $c_p$, we obtain the effective  action for the  anharmonic cavity, \begin{equation}
   i\mathbf{S}_{\rm ac}[\bar a,a,\phi]=i\int \bar a(t)\check {\mathcal{G}}_{\rm ac}^{-1}[\phi(t)]a(t) dt \ ,
   \end{equation} 
   where 
   the    inverted Green function
   \begin{equation}
   \check {\mathcal{G}}_{\rm ac}^{-1}[\phi] = \delta(t - t') \check G_{\rm ac}^{-1} [\phi(t')]+ \check \Sigma(t-t')  \label{G-ao}
   \end{equation}
    becomes   a nonlocal in time   
   due to the self-energy   $\check \Sigma(t)=\int\frac{d\omega}{2\pi}e^{-i\omega t}\check \Sigma_\omega$. It has the causality structure,  $
   \check \Sigma_\omega  =  
   \begin{bmatrix}
   0 &&  \Sigma^{A}_\omega
   \\   
   \Sigma^{R}_\omega  && \Sigma^{K}_\omega
   \end{bmatrix} 
   $, the retarded and advanced components are $\Sigma^{R(A)}_\omega =\pm i \Gamma$ where the total  relaxation $\Gamma= \Gamma_{\rm L}+\Gamma_{\rm R} $ is a sum of the rates   determined by the    couplings  to left and right waveguides, $\Gamma_{\rm L,R}=\pi \nu_{\rm L,R} |t_{\rm L,R}|^2$. The  Keldysh component, $\Sigma^{K}_\omega = 2 i  \Gamma (2N_{{\rm ac},\omega}+1)$,  brings   information on the nonequilibrium distribution function in the  cavity mode  
       $
    N_{{\rm ac},\omega} =  \frac{\Gamma_{\rm L}}{\Gamma}N_{{\rm L}, \omega} +  \frac{\Gamma_{\rm R}}{\Gamma}N_{{\rm R}, \omega} $.   
   
   The integration over the cavity fields  $\bar a , a$ provides the effective action for the fluctuating potential  \begin{equation}i\mathbf{S}_{\rm eff}[\phi_c,\phi_q] = \int \frac{i}{  \epsilon} \phi_c \phi_q  dt-{\rm Tr} \ln \big(\check {\mathcal{G}}_{\rm ac}^{-1}[\phi_c,\phi_q]\big) \ . \label{S-eff}
   \end{equation}   The trace and logarithm here are determined over discretized time axis. As long as this action  is non-Gaussian due to the matrix logarithm, we apply Gaussian expansion near a saddle point.

 \subsubsection*{Saddle point equation
}
A configuration of $\phi_c(t)$ that determines a saddle point   of  $\mathbf{S}_{\rm eff}$ is  given by the   zero variation with respect to $\phi_q$, $\frac{\delta}{\delta\phi_q}\mathbf{S}_{\rm eff}[\phi_c,\phi_q]=0$. 
Applying this to  (\ref{S-eff}), we find $ \int  \phi_c  dt  -i  \epsilon {\rm Tr}    \check {\mathcal{G}}_{\rm ac} [\phi_c] =0$. Here, we use the standard notation for trace ${\rm Tr} (f(t-t'))=f(0) \int  dt = \int \frac{d\omega}{2\pi} f_\omega \int  dt  $, where the time integral stands for a large constant that cancels out in  the saddle point equation.  
In our system, the saddle point configuration is associated with the static part of the field, $\phi_0$.  An explicit form of   this equation  is
\begin{equation}   \phi_0  - \epsilon-\frac{\epsilon \Gamma}{\pi} \int \frac{N_{{\rm ac},\omega} d\omega}{(\omega_0-\epsilon+\phi_0)^2+\Gamma^2} =0 \ .  \label{phi0}
\end{equation} 
In our case of flat $N_{{\rm L},\omega}= F  $ in the range $\omega\in [\omega_0-\Delta; \ \omega_0+\Delta]$ with $\Delta\gg \Gamma, \epsilon$, and the absence of incident modes in the right waveguide,  {\it i.e.}, $N_{{\rm R},\omega}=0$, the integral in (\ref{phi0})  gives a constant  that does not depend on $\phi_0$. The equation  (\ref{phi0}) becomes linear and we find
    $ 
    \phi_0= \epsilon +F\epsilon\frac{\Gamma_{\rm L}}{\Gamma} $.
    
    Thus, the photon Green function at the saddle point    is found after the inversion of $\check {\mathcal{G}}_{\rm ac}^{-1}[\phi]$ from (\ref{G-ao}) with   $\phi=\phi_0$:
\begin{equation}
 \check   {\mathcal{G}}_{{\rm ac},\omega}[\phi_0] =  
	\begin{bmatrix}
		\mathcal{G}_{{\rm ac},\omega}^K &&  	\mathcal{G}_{{\rm ac},\omega}^R
		\\ \\ 
			\mathcal{G}_{{\rm ac},\omega}^A  && 0
	\end{bmatrix}\ , \quad \mathcal{G}_{{\rm ac},\omega}^{R(A)}=\frac{1}{\omega-
	(\omega_0-\epsilon+\phi_0)\pm i\Gamma} \ , \quad \mathcal{G}_{{\rm ac},\omega}^K =\frac{-2i\Gamma (N_{{\rm ac},\omega}+1)}{(\omega-(\omega_0-\epsilon+\phi_0)
	)^2+\Gamma^2} \ . \label{G-ao-1}
\end{equation}

    \subsubsection*{Caldeira-Leggett approach for fluctuations } 
    \label{Results:Quasi-classical}
     Non-stationary configurations of the field, $\delta\phi_c(t)=\phi_c(t)-\phi_0$,  determine    many-body transitions which induce decoherence. 
Let us go back to the effective action $\mathbf{S}_{\rm eff}$. It is non-Gaussian because of the matrix logarithm. 
We apply    quasi-classical  second-order expansion of the logarithm by stochastic, $\delta\phi_c$, and quantum, $\phi_q$, fluctuations:
\begin{multline}
	{\rm Tr} \ln \big(\check {\mathcal{G}}_{\rm ac}^{-1}[\phi(t)
	]\big) = 	{\rm Tr} \ln \big(\check {\mathcal{G}}_{\rm ac}^{-1}[\phi_0
	]-\delta\phi_c \check \tau_x -\phi_q\check \tau_0\big)  
	\approx {\rm Tr} \ln \big(\check {\mathcal{G}}_{\rm ac}^{-1}[\phi_0
	]\big)-\frac{1}{2}{\rm Tr}\left(\check {\mathcal{G}}_{\rm ac} [\phi_0
	] \big(\delta\phi_c \check \tau_x+ \phi_q\check \tau_0\big)
	\right)^2
\ .  \label{TrLn} 
\end{multline}
This approximation for  $\mathbf{S}_{\rm eff}$ provides Caldeira-Leggett     action   (\ref{S-eff-1})  
which rules fluctuational and dissipative dynamics of $\phi(t)$.
For nonequilibrium  distributions indicated above, we find from (\ref{TrLn}) that   the retarded and advanced parts in (\ref{S-eff-1}) read     
 \begin{equation}
	\alpha^{R(A)}_\omega  = \frac{\pm 16i}{\pi\Gamma\Delta^3} \epsilon F^2 \Gamma_{\rm L}^2  + O[\Delta^{-4}]\ .
\end{equation}
	 They are not universal, {\it i.e.}, they depend on the bandwidth and vanish as $1/\Delta^{ 3}$. The Keldysh part, however, is non-zero  
 \begin{equation}
\alpha^K_\omega  = 16 \frac{ \Gamma_{\rm L} F (\Gamma +\Gamma_{\rm L} F)}{\Gamma  \left(4 \Gamma ^2+\omega ^2\right)} \ . \label{alphaK}
\end{equation}
This fact demonstrates a break of  fluctuation-dissipation relation out of the  equilibrium, {\it i.e.},  $ \alpha^K_\omega\neq(2N_{{\rm ac},\omega}+1)
\left(\alpha^R_\omega - \alpha^A_\omega\right) $.

We note that the   action $\mathbf{S}_{\rm CL}  $ determines stochastic Langevin equation 
$\left(\frac{1}{\epsilon}+\frac{1}{2}\alpha^{R}_\omega\right)\delta\phi_{c,\omega}=\xi_\omega
$
 where the stochastic force $\xi$ has the correlator   given by   $\langle \xi_{-\omega}\xi_\omega\rangle=\frac{1}{4}\alpha^K_\omega$.  
Addressing the  photons decoherence in Keldysh formalism, we introduce  stochastic  $\Phi(t)=\int\limits^t_{-\infty}\delta\phi_c(t')dt'$ and quantum $\varphi(t)=\int\limits^t_{-\infty} \phi_q(t')dt'$ phases. In our case with vanishing $\alpha^R$, the Langevin equation for the transmitted photon phase is reduced to 
\begin{equation}\frac{d}{dt} \Phi(t)=  \epsilon \xi(t) \label{LE}
\end{equation} that corresponds to a frictionless drift.

\subsection*{Calculations of transmission functions}
\subsubsection*{
Non-interacting limit. Lorentzian spectrum.	Analogy with Landauer formula  }
\label{Results:Landauer}

   We address a photon transport  in  a spirit of the Landauer formula determined by the transmission probability $T_\omega$. We start from a   single-particle problem at     $\epsilon=0$.  This limit is  exactly solvable.  After that, we study the case of $\epsilon\neq 0$   taking into account many-body interaction utilizing  Caldeira-Leggett action (\ref{S-eff-1}).

Let us define the photon current as time derivative of the photon number operator $\hat n_{\rm R}= \sum_p \hat c_p^\dagger \hat c_p$ in the right waveguide as  $J=\frac{d}{dt}\langle \hat n_{\rm R}\rangle=\frac{i}{\hbar}\langle[\hat H, \hat n_{\rm R}]\rangle $. 
	 	 We note, that   tunnelling Hamiltonians can be represented as  $\hat H_{\rm tL}=t_{\rm L} \hat {{B}} ^ \dagger  \hat a+ t_{\rm L}^*  \hat a ^ \dagger  \hat {{B}} 
$ and $ \hat H_{\rm tR}=t_{\rm R}   \hat {C} ^ \dagger  \hat a+t_{\rm R}^* \hat a^ \dagger \hat {C}
$ where $\hat {{B}}=\sum_k  \hat b_k $ and $\hat {C}=\sum_p  \hat c_p$ are 'local' operators of waveguide   fields. In  this representation, the photon current reads as
   \begin{equation}
J= i t_{\rm R}^*\langle  \hat a^\dagger\hat {C}  \rangle  - i t_{\rm R}\langle  \hat {C}^\dagger \hat a \rangle \ . \label{J}
   \end{equation}
   We employ Keldysh Green functions technique to calculate these averages.  
      In the non-interacting case,  the   action      reads
       \begin{equation} 
    	i\mathbf{S}^{(0)} 
    	=  i\! \int \frac{d\omega}{2\pi} 
    	\bar\psi_\omega \mathbb{G}_\omega^{-1} \psi_\omega \ , \quad 
    	\mathbb{G}_\omega^{-1} =\begin{bmatrix}
    		\check	 g_{\rm L,\omega}^{-1} && -t_{\rm L}\check\tau_x && 0 \\ \\ 
    		-t_{\rm L}^*\check\tau_x  && \check G_{0,\omega}^{-1}  && -t_{\rm R}^*\tau_x  \\ \\ 
    		0 && - t_{\rm R}\check \tau_x && \check g_{\rm R,\omega}^{-1} 
    	\end{bmatrix} \ , \quad 
    	    	 	\check 	g_{\rm L,R,\omega}^{-1} =    
    \begin{bmatrix}
    0 && -i\frac{1}{\pi \nu_{\rm L,R}} 
    \\ \\ 
    i\frac{1}{\pi \nu_{\rm  L,R}} && \frac{2i}{\pi \nu_{\rm L}}(2N_{{\rm L,R},\omega}+1)
\end{bmatrix}
    	\ .  \label{S0}  
    \end{equation}
The  inverted Green functions of the cavity mode is
$
	\check G_{0,\omega}^{-1}= (\omega-\omega_0)   \check \tau^x
	 $. Green functions of   left and right waveguides   are, respectively,    $ 	\check g_{\rm L, \omega}=-  \sum_k\check	 G_{{\rm L},\omega,k} $  and $\check	 g_{\rm R,\omega} = - \sum_p\check	 G_{{\rm R,\omega},p}  $.  They act in space of  variables $\psi=[ {B}_\omega, \ a_\omega, \  {C}_\omega]$. The summations over $k$ and $p$   are found after replacements  $\sum_{k,p} \to \nu_{\rm L,R}\int dE$.

Considering single-particle action $\mathbf{S}^{(0)}$,     distribution functions $N_{{\rm L,R},\omega}$   can be arbitrary.   Inverting $\mathbb{G}_\omega^{-1}$ from (\ref{S0}), we find   cavity-to-right waveguide propagator
    \begin{equation}
   	\check g_{0\rm R,\omega} = -2i\pi  \nu_{\rm R}t^*_{\rm R}	
	\begin{bmatrix}
   		\frac{ (\omega - \omega_0)( 2N_{\rm R,\omega} +1) - 2i\Gamma_{\rm L} (N_{\rm R,\omega} -N_{\rm L,\omega}) }{ \Gamma^2 +(\omega - \omega_0)^2} &&   \frac{
   			1/2}{ \omega - \omega_0 +i \Gamma }  
   		\\ \\ 
   		\frac{
   			-1/2}{ \omega - \omega_0 -i \Gamma }   && 0
   	\end{bmatrix}  \ . \label{G0R}
   \end{equation}
   We use these expressions in the current (\ref{J}): $\langle  \hat {C} ^\dagger\hat a \rangle=\int i g^{<
   }_{0\rm R, \omega}  \frac{d\omega}{2\pi} $ and  $\langle   \hat a^\dagger\hat {C} \rangle=\langle  \hat {C} ^\dagger\hat a \rangle^*
    $ where   the  'greater' Green function $g^{<  }_{0\rm R, \omega}     =  \frac{1}{2}\left(g^K_{0\rm R, \omega} -g^R_{0\rm R, \omega} +g^A_{0\rm R, \omega} \right)  $ 
    is introduced.
   As a result, we arrive at the  Landauer formula for the photon current
       \begin{equation}
   	J=\int\tau_{  \omega}(N_{\rm L,\omega} -N_{\rm R,\omega} ) \frac{d\omega}{2\pi} \ .  \label{Landauer}
   \end{equation}
 The transmission probability  has   the Lorentzian form
  $ 
   	\tau_{  \omega} = 
   	\frac{4\Gamma_{\rm L}\Gamma_{\rm R}}{(\omega - \omega_0)^2+\Gamma^2 }\  
   	 $.  
It reproduces unitary transmission $\tau_{\omega{=} \omega_0}=1$ at the resonance  and symmetric relaxation rates $\Gamma_{\rm L}=\Gamma_{\rm R}=\Gamma/2$, the well-known result of the input-output theory~\cite{Gardiner:2004aa}.

\subsubsection*{Interacting case. 
Gauge transformation }
\label{Results:anharmonic}
In this part, we present a generalization on the many-body problem where $\epsilon\neq 0$ addressing    the situation of flat distribution functions $N_{{\rm L} ,\omega }=F$ and $N_{{\rm R},\omega }=0$. 
 The  action   with stochastic potential becomes local in time  due to flat distributions
\begin{equation}
	i\mathbf{S} 
	=  i\! \int dt 
	\begin{bmatrix} \bar{B}_t & \bar a_t & \bar {C}_t
	\end{bmatrix} 
	\begin{bmatrix}
		\check	 g_{\rm L}^{-1}  && -t_{\rm L}\check\tau_x && 0 \\ \\ 
		-t_{\rm L}^*\check\tau_x  && \check G_{\rm ac}^{-1}[\phi]  && -t_{\rm R}^*\tau_x  \\ \\ 
		0 && - t_{\rm R}\check \tau_x && \check g_{\rm R}^{-1}  
	\end{bmatrix} 
	\begin{bmatrix} {B}_t \\  \\ a_t \\  \\ {C}_t \label{Seps}
	\end{bmatrix}
	\ .
\end{equation}
In contrast to the action in the non-interacting limit $\mathbf{S}^{(0)}$, the Green function $\check G_0^{-1}$ in the interacting case   is replaced with $ \check G_{\rm ac}^{-1}[\phi]=(i\partial_t -\omega_0+\epsilon-\phi_0-\delta\phi_c(t))\check \tau_x- \delta\phi_q(t) \check \tau_0$.   
The locality   in time gives an advantage in analytic calculations. Namely, it is possible to get rid of time-dependent $\delta\phi_c(t)$ and $ \phi_q(t)$ by the gauge transformation. This transformation from $\psi = [B, \ a, \ C]^T$ to new fields, $\psi'= [B', \ a', \ C']^T$, is  distinct on upward and backward parts of $\mathcal{K}$ and reads
 \begin{equation}
 \psi_\pm(t)= e^{-i \Phi(t)\mp i \varphi(t)} \psi'_\pm(t)
\ . \label{gauge}
 	 \end{equation}
 It provides a nonperturbative solution for    photon decoherence. A dynamics of these  new fields $\psi'$ is ruled by the   action $\mathbf{S}^{(0)}$ found for $\epsilon=0$. Hence, the solution for   cavity-to-right   waveguide propagator $\check g_{\rm ac R}$ at $\epsilon\neq 0$   is obtained from  non-interacting result $\check g_{\rm 0R}$ from (\ref{G0R})    when  the gauge inversed to (\ref{gauge})  is applied:
\begin{equation}
\check g_{{\rm ac R}
}(t,t') = \Big(\cos\varphi(t)\cos\varphi(t')- i \sin(\varphi(t) + \varphi(t'))\Big) e^{-i(\Phi(t)-\Phi(t'))}e^{i(\epsilon-\phi_0)(t-t')} \check g_{\rm 0R}(t-t') .  \label{g-tt}
\end{equation}
This function is not stationary due to the fluctuating potential.  After the  averaging $\check g_{{\rm ac R} 
}(t,t')$ with respect to $\mathbf{S}_{\rm CL}$, we arrive at the stationary Green function  that depends on the time difference
\begin{equation}
	\langle\check g_{{\rm ac R}
	}(t, t') \rangle = z(t-t') e^{-i\langle\Phi(t)\varphi(t')\rangle+i\langle\varphi(t)\Phi(t')\rangle} 
	\check g_{\rm 0R}(t-t') \ . \label{g-tt-av}
	\end{equation} 
	The relevant modification appears in  the envelope related to stochastic fluctuations $z(t)=\langle e^{-i(\Phi(t)-\Phi(0))}\rangle$. The average with Gaussian action  $\mathbf{S}_{\rm CL}$ yields 	$
z(t)=\exp(-D(t))$ where  the  symmetrized  autocorrelation function is $D(t )=\frac{1}{2} (\langle\Phi(t) \Phi( 0)\rangle + \langle\Phi(0) \Phi( t)\rangle ) -  \langle\Phi^2(0)\rangle $. 
  With the use of Langevin equation (\ref{LE}), we find that $ D(t)=\frac{1}{4}\int^t_0\int^{t'}_0\alpha^K(t'')dt''dt'$. The integration with Keldysh component $\alpha^K (t)=   F(F+2) e^{-2 \Gamma  |t|}$, found after the Fourier transform of (\ref{alphaK}),  gives
   \begin{equation}
	D(t)=  
	 \frac{ \Gamma_{\rm L}  \epsilon ^2 }{2 \Gamma ^3}\left( F +\frac{\Gamma_{\rm L}}{\Gamma} F^2 \right) \left(2 \Gamma  |t|  +e^{-2 \Gamma  |t|}-1\right) \ . \label{D}
\end{equation}  
 Combinations of classical-quantum retarded and advanced correlators in (\ref{g-tt-av}) are due to the mixing of   fields on different parts of the contour $\mathcal{K}$; they read  $\langle\Phi(t)\varphi(0)\rangle=i\epsilon t \theta(t)$ and $\langle\varphi(t)\Phi(0)\rangle=-i\epsilon t \theta(-t)$.  In our Gaussian theory, classical-quantum terms yield  simply a complex phase $e^{-i\epsilon (t-t')}$ that results in a shift of the cavity mode frequency    
\begin{equation} 
	\omega_{\rm ac}= \omega_0 -\epsilon +F\epsilon \frac{\Gamma_{\rm L}}{\Gamma}  \ , \label{ao}
\end{equation}  a quantum counterpart of  the  nonlinear frequency shift in a classical anharmonic oscillator.

   \subsubsection*{Gaussian spectrum }
    The result (\ref{g-tt-av})  provides the expression for the transmission function  in the time domain
     \begin{equation}
      	T(t)=z(t)   \tau(t) \ , \quad  \tau(t) =2 \frac{\Gamma_{\rm L} \Gamma_{\rm R}}{  \Gamma} e^{-\Gamma  |t| - i\omega_{\rm ac} t} \ . \label{T-t}
      	 \end{equation}
       Here,  $ \tau(t)$ is Fourier transformed   Lorentzian transmission  with the maximum at $\omega_{\rm ac}$.
                      The Fourier transform of (\ref{T-t}) reads $
       	T_\omega=\int z_{\omega-\omega'}   \tau_{\omega'} \frac{d \omega'}{2\pi}  $.
         Analytic  calculation  of this integral  is challenging. Nevertheless, we find asymptotic expressions   for the strongly nonequilibrium regime  with the use of the saddle point method. In this solution, the correlator in the exponent of  $z(t)$ is approximated as $D(t)\approx  \kappa^2 t^2$. Here, new relaxation rate  $
      	\kappa= \epsilon  \sqrt{ \frac{\Gamma_{\rm L}   }{\Gamma } \left(F+\frac{\Gamma_{\rm L}}{\Gamma} F^2\right)}
      	$  appears. 
    It shows that the coherence loss rate grows non-linearly  as    $F$ increases.  
      The   Fourier transformation gives the   transmission spectrum of  Gaussian shape:
  	$
	T_{\omega} = 2\frac{\sqrt{\pi}\Gamma_{\rm L}\Gamma_{\rm R}}{ \Gamma\kappa} \exp\left(\!-\frac{(\omega-\omega_{\rm ac})^2}{4\kappa^2} \right)  $.
    This result applies to  a regime of strong driving or strong interaction, such that the condition $\kappa\gg \Gamma$ is satisfied.  Let us analyze this condition for the symmetric  case of    $\Gamma_{\rm L}=\Gamma_{\rm R}= \Gamma/2$.  The  decay rate can be equivalently expressed through the current as $
  	\kappa = \frac{\epsilon}{\Gamma}\sqrt{J\Gamma+J^2} $ where  $J=F\Gamma/2$  does not depend on anharmonicity in our problem with the  wide spectrum of the incoherent drive.
The   condition	$\kappa\gg \Gamma$ is resolved as $J\gg J^*_\epsilon$:
  \begin{equation}
  J^*_\epsilon=\frac{\Gamma}{2}\Big(\sqrt{1+4\frac{\Gamma^2}{\epsilon^2}}-1 \Big)\approx  \begin{cases} \frac{\Gamma^2}{\epsilon}  \ , & \epsilon\ll\Gamma \ ; \\ \\
   \frac{\Gamma^3}{\epsilon^2}  \ , & \epsilon\gg\Gamma \ . \\
  \end{cases} \label{J-cond}
  	 \end{equation}   
The scale $J^*_\epsilon$  corresponds to the lower boundary for the current at a given $\epsilon$ when it drives the system  out of  the equilibrium   and induces  Gaussian chaotic light emission. If the current is small, $J\ll J^*_\epsilon$, then the system is in a fixed point related to thermal behavior, as shown in the phase diagram in Fig.~\ref{diagramnoise}.
We learn from (\ref{J-cond}) that there is two asymptotics  in $J^*_\epsilon$ that distinguish weak, $\epsilon\ll\Gamma$, and strong, $\epsilon\gg\Gamma$, interaction limits. Both of them overlap with the Lorentzian and Gaussian sectors.

	\subsubsection*{Photon current and cavity photon number }
	
As it follows from  (\ref{Landauer}), the photon  current at   $N_{{\rm L}, \omega}=F$ and $N_{{\rm R}, \omega}=0$	 reads 
	\begin{equation} 
		J=2 F \frac{\Gamma_{\rm L}\Gamma_{\rm R}}{\Gamma}
		\ . \label{J-0}
	\end{equation}  
	The average photon number in this limit,
	\begin{equation} 
		\langle\hat a^\dagger \hat a\rangle  =F \frac{ \Gamma_{\rm L} }{ \Gamma}     \ , \label{n}
	\end{equation} 
 is obtained as $  \langle\hat a^\dagger \hat a\rangle=\int i\mathcal{G}_{{\rm ac},\omega}^< \frac{d\omega}{2\pi}
	$ where the  'greater' Green function is $\mathcal{G}_{{\rm ac},\omega}^< = \frac{1}{2}(\mathcal{G}_{{\rm ac},\omega}^K -\mathcal{G}_{{\rm ac},\omega}^R +\mathcal{G}_{{\rm ac},\omega}^A )$.  	We note that  the   frequency shift due to the anharmonicity  can be represented via (\ref{n}) as $\omega_{\rm ac}=\omega_0 - \epsilon+\langle\hat a^\dagger \hat a\rangle \epsilon$. As it should be, the same value can be found from  a mean-field approach for Bose-Hubbard interaction   where    $\hat a^\dagger\hat a \hat a^\dagger\hat a \to  \hat a^\dagger\hat a \langle\hat a^\dagger \hat a\rangle $. 
	
	The current given by (\ref{J-0}) is not modified by the interaction   in the case of flat distribution functions. As follows from the representation of the current through  $\langle  g_{{\rm ac R} }^<(t, t) \rangle $  at coincident times, we find $J=F z(0) \int\frac{d\omega}{2\pi}\tau_{\omega}$ from (\ref{T-t}). As long as  $z(0)=1$ for any interaction, $J$ does  not depend on $\epsilon$. The same logic applies to   $\langle\hat a^\dagger \hat a\rangle $  which also does  not change as $\epsilon$ increases.

   \subsection*{Calculations of the noise and intensity  fluctuations} 
   \label{Results:noise}
   \subsubsection*{Generating functional method}
 We start calculations of noise from the non-interacting limit and then generalize for the interacting case. Cumulants of transmitted photons can be calculated through variations of the generating functional logarithm, $  \ln Z[\eta]$.
The generating functional is given by the path integral  
     \begin{equation}
     	Z[\eta]=\int D[\bar \psi, \psi]\mathcal{T}\exp(i\mathbf{S}[\bar\psi, \psi]+i \eta \bar\psi \mathbb{J} \psi) \ , \quad \mathbb{J} = \begin{bmatrix}
     		0  & 0 & 0 & 0  & 0 & 0 \\ 
0  & 0 & 0 & 0  & 0 & 0 \\ 
0  & 0 & 0 & 0  & 0 & 0 \\ 
0  & 0 & 0 & 0  & i t_{\rm R}^* & 0 \\ 
0  & 0 & 0 & 0  & 0 & 0 \\ 
0  & 0 & -it_{\rm R} & 0  & 0 & 0 \\ 
     	\end{bmatrix} \ , \quad \psi(t) =\begin{bmatrix}
     	 B_+(t) \\ 
     	 B_-(t) \\ 
     	 a_+(t) \\ 
     	 a_-(t)\\ 
     	C_+(t) \\ 
     	C_- (t)\\ 
     \end{bmatrix} \ .  \label{S-eta}
     \end{equation} Here,  $  \eta$ is generating variable and  $\mathbb{J}$   parametrizes current  from (\ref{J})  through fields $a $ and ${C} $ on opposite branches (labeled by $\tau_z=\pm1$) of the contour $\mathcal{K}$ as  $J=\bar\psi \mathbb{J} \psi$. The integration over $\bar \psi,   \psi$ in (\ref{S-eta}) results in $\ln Z[\eta]={\rm Tr}\ln (1+\mathbb{G}\eta   \mathbb{J})$. Photon current $J$ is found as first variation of $\ln Z[\eta]$  by $\eta$, where  it is set to zero   $\eta\to 0$. This gives $J=\int\frac{d\omega}{2\pi}i {\rm tr}\big( \mathbb{G}_\omega\mathbb{J}\big)$ that is reduced to Landauer formula (\ref{Landauer}); here, "$\rm tr$" stands for trace over $\check\tau$, $L$, $R$, and oscillator indices.  In order to find  the spectrum of the  symmetrized noise,  $S	^{(0)}_\omega=\frac{1}{2} (\langle\delta J_{-\omega} \delta J_{\omega} \rangle+\langle\delta J_{\omega} \delta J_{-\omega} \rangle)$,  we need to find second variation of $\ln Z[\eta]$ by $\eta_{-\omega}$ and $\eta_{\omega}$. In our particular case of flat $N_{{\rm L},\omega}$, the spectrum is Lorentzian 
   \begin{equation}
   S	^{(0)}_\omega = -\frac{1}{2}\int\frac{d\omega_1}{2\pi}i {\rm tr}\big( \mathbb{G}_{\omega_1}\mathbb{J} \mathbb{G}_{\omega+\omega_1}\mathbb{J}\big) 
   =8F^2\frac{\Gamma_{\rm L}^2\Gamma_{\rm R}^2}{\Gamma(\omega^2+4\Gamma^2)}  \ . \label{S-0}
   \end{equation}

     \subsubsection*{Asymptotic behavior of the noise-current relation}
    Hereafter, we express $F$ through the current $J$ according to (\ref{J-0}) and suppose that $\Gamma_{\rm L}$ and $\Gamma_{\rm R}$ can be unequal. This allows   analyzing zero-frequency noise-current ratio, $S(J)$, where both $S$ and $J$ are measured by the output detector.
      The zero-frequency noise found in (\ref{S-0}) has quadratic scaling,     $
   S_{\rm therm}
   =  \frac{J^2}{2\Gamma}  $ at $   J\ll J^*_\epsilon $ , 
   that corresponds to Lorentzian light emitted into the output waveguide.

  The decoherence  is important at $J\gg J^*_\epsilon$ when pseudothermal output  light  becomes  Gaussian and the thermal noise is changed to   shot noise.
Technically, the inclusion of the stochastic $\Phi(t)$ in the noise calculation  is performed with the gauge transformation, similarly to that applied  for the transmission function $T(t)$. Symmetrized correlator of the noise, $S(t)=\frac{1}{2}(\langle\delta J(t)\delta J(0)\rangle + \langle\delta J(0)\delta J(t)\rangle)$, reads
$S(t)= z^2(t) S^{(0)}(t)$ where  $ S^{(0)}(t) = \frac{1}{2} J^2e^{-2\Gamma |t|} $  is  time-resolved  Lorentzian correlator found in the non-interacting limit (\ref{S-0}). 
The shot noise expression follows from the Fourier transform  of $S(t)$ where the correlator in the exponent of  $z(t)$ is Gaussian $D(t)\approx  \kappa^2 t^2$. The   low-frequency   result, $S\equiv S_{\omega=0}$,  is given by   $S=J^2\int\limits_0^\infty  e^{-2\Gamma |t|-2\kappa^2 t^2} $. Assuming $\kappa\gg \Gamma$, that is equivalent to  $J\gg J^*_\epsilon$, we calculate this integral and  arrive at one of central results
\begin{equation}
S = \sqrt{\frac{\pi}{2}} \frac{\Gamma_{\rm R}}{ \epsilon} \frac{J}{\sqrt{1+2\Gamma_{\rm R}/J}}\ .\label{S}
\end{equation}
Let us analyze it in details. 
If the interaction is weak, $\epsilon\ll \Gamma$, then we always have $\Gamma_{\rm R}/J\ll 1 $  according to   $J^*_\epsilon \sim \frac{\Gamma^2}{\epsilon} $ from (\ref{J-cond}). 
Thus, in the nonequilibrium regime at weak anharmonicity, we find conventional shot noise from (\ref{S})  with a linear noise-current ratio
$
S_{\rm shot} = \sqrt{\frac{\pi}{2}} \frac{\Gamma_{\rm R}}{ \epsilon}  J  $ at $ J\gg 
\frac{\Gamma^2}{\epsilon}$. 
 This result is considered as a nonequilibrium fixed point  with  the exponent  $\gamma=1$.
 
For strong interaction, $\epsilon\gg\Gamma_{\rm L,R}$, the thermal noise sector in Fig.~\ref{diagramnoise}  shrinks because of the vanishing $J^*_\epsilon\sim\frac{\Gamma^3}{\epsilon^2}$. This means that we can go to low current   domain where $J\ll \Gamma_{\rm R}$. Here,   the noise-current relation  (\ref{S}) has the following asymptotic
\begin{equation}
S_{\rm shot}' = \frac{\sqrt{\pi\Gamma_{\rm R}}}{2\epsilon }   J ^{3/2} \ , \quad  \Gamma\gg  J\gg \frac{\Gamma^3}{\epsilon^2}\ .\label{S-crit}
\end{equation}  
This non-analytical dependence occurs at small $J\sim J^*_\epsilon$, {\it i.e.}, it is parametrically close to zero current line  $J=0$  in Fig.~\ref{diagramnoise} at   large $\epsilon$. This feature is also demonstrated in Fig.~\ref{exponent} for $\epsilon>\Gamma$ where orange and red curves drop rapidly from $\gamma=2$ to $\gamma=3/2$. Then, curves saturate to conventional shot noise   with $\gamma=1$ as the current increases.

There are smooth crossovers between  $S_{\rm therm} $, $S_{\rm shot}'$  and $S_{\rm shot} $ at intermediate $\epsilon\sim\Gamma$ as a consequence of     fluctuations in  zero-dimensional cavity. 
At weak anharmonicity,  there is no  $S_{\rm shot}'$ behavior. Instead, there is a crossover from $S_{\rm therm} $ to $S_{\rm shot} $ at $J
\sim  \frac{\Gamma^2}{\epsilon}$. 
It is shown in Fig.~\ref{exponent}  where blue curve decays smoothly from $\gamma=2$ to $\gamma=1$.

\subsubsection*{Intensity correlators
}

Time-resolved intensity-intensity correlator in the right waveguide, \begin{equation}
	g^{(2)}(t)=\frac{\langle  \hat C^\dagger\! (t) \hat C(t) \hat C^\dagger\! (0) \hat C(0)\rangle}{|\langle \hat C^\dagger\! (0) \hat C(0)\rangle|^2} \ , \end{equation}
is an indicator for bunching or antibunching of photons in the output field. In our solution we assumed that anomalous term $\langle   \hat C(t)  \hat C(0)\rangle$ is zero, and this four-point correlator is  defined through the two-point one, 
$	g^{(2)}(t)=1+|g^{(1)}(t)|^2 $, according to the Wick theorem applicable in our quasi-classical solution. We find $g^{(1)}(t)=z(t)\check g_{\rm RR}(t )/\check g_{\rm RR}(0 )$ where  $  \check g_{\rm RR}(t )\propto e^{-\Gamma t}$ is the Lorentzian propagator      while $z(t)$ involves $\epsilon$ and $F$.  

Time dependence of the correlator shows exponential decay in the non-interacting limit  $g^{(2)}(t)=1+e^{-2\Gamma t}$ at  $t>0$.  In the interacting case,  we find
\begin{equation}
	g^{(2)}(t)=1+ {\rm exp}\Big[-
	\frac{\kappa^2}{\Gamma^2}(2\Gamma t +  e^{-2 \Gamma  |t|}-1  )-2\Gamma t\Big]  \ .
\end{equation}
There is Gaussian law of the correlations decay  $g^{(2)}(t)=1+e^{-2 \kappa^2  t^2}$  for $t\ll 1/\Gamma$. For large timescale, $t\gg 1/\Gamma$, there is  exponential decay   $g^{(2)}(t)=1+e^{-2\frac{\kappa^2}{\Gamma } t}$. 
As follows from the condition $\kappa\gg\Gamma$ on the nonequilibrium, the decay rate in the latter case   exceeds that from the non-interacting limit as       $\frac{\kappa^2}{\Gamma } \gg \Gamma$.

 \section*{Acknowledgements}

D.S.S.  is grateful to Walter V. Pogosov, Oleg V. Astafiev,  Arkady M. Satanin,   Evgeny S. Andrianov, and Alexander Shnirman for helpful discussions. The major part of the reported study was funded by RFBR according to the research project N\textsuperscript{\underline{o}}    
19-32-80014. We also acknowledge the financial support  by RFBR according to  research projects N\textsuperscript{\underline{o}}~20-37-70028 and N\textsuperscript{\underline{o}}~20-52-12034. The work reported in Section "Calculations of transmission functions" was financed exclusively by the Russian Science Foundation under Grant N\textsuperscript{\underline{o}}  16-12-00095. 

\section*{Author contributions statement}

D.S.S. derived and analyzed all reported results, reviewed and wrote the final version of  the manuscript.

\section*{Additional Information}
Competing Interests Statement: The author declare no competing interests.

\end{document}